\documentclass[a4paper,11pt]{article}
\pdfoutput=1

\usepackage[symbol]{footmisc}
\usepackage{jheppub}

\usepackage{rotating}
\usepackage{tikz}
\usepackage{makecell}
\usepackage{float}
\usepackage{environ}
\usetikzlibrary{decorations.pathmorphing,decorations.markings,trees,shapes}
\usepackage[force]{feynmp-auto}
\tikzset{graviton/.style={
    decoration={coil, aspect=0, amplitude=2pt, segment length=3pt}, 
    decorate,
    double
  },
    photon/.style={decorate, decoration={snake}},
    electron/.style={draw=black, postaction={decorate},decoration={markings,mark=at position .55 with {\arrow[draw=black]{>}}}},
        antielectron/.style={draw=black, postaction={decorate},decoration={markings,mark=at position .55 with {\arrow[draw=black]{<}}}},
    gluon/.style={decorate, draw=magenta,
        decoration={coil,amplitude=4pt, segment length=5pt}} 
}
\usepackage{amsfonts}
\usepackage{amssymb}
\usepackage{amsmath}
\usepackage{amssymb}
\usepackage{wasysym}
\usepackage{bm}
\usepackage{latexsym}
\usepackage{textcomp}
\usepackage{subcaption}
\usepackage{physics}

\usepackage{verbatim}
\usepackage{slashed}
\usepackage{array}
\usepackage{graphicx}
\usepackage{multirow}
\usepackage{graphicx} 
\usepackage{color}
\usepackage{dsfont}

\usepackage{tikz-feynman}
\usepackage{feynmp-auto}
\usepackage{amssymb,amsmath,bm}
\usepackage{color}
\usepackage{slashed}
\usepackage{stackengine}
\usepackage{hyperref}
\usepackage{relsize}
\usepackage{graphicx}
\usepackage[T1]{fontenc}
\usepackage[utf8]{inputenc}
\usepackage{amsfonts}
\usepackage{pifont}
\usepackage{amsthm}
\usepackage{booktabs}
\usepackage{array}
\usepackage{longtable}
\usepackage{lscape} 
\usepackage{rotating}
\usepackage{float}
\usepackage{multirow}
\usepackage{graphicx,rotating,amsmath,multirow}
\usepackage{hyperref,slashed}
\usepackage{upgreek}

\usepackage{pdflscape} 

\usepackage{fontawesome} 

\usepackage[detect-all]{siunitx} 
\usepackage{mathrsfs}            

\graphicspath{{img/}}

\definecolor{rosso}{cmyk}{0,1,1,0.4}
\definecolor{rossos}{cmyk}{0,1,1,0.55}
\definecolor{rossoc}{cmyk}{0,1,1,0.2}
\definecolor{blu}{cmyk}{1,1,0,0.3}
\definecolor{blus}{cmyk}{1,1,0,0.6}
\definecolor{bluc}{cmyk}{1,1,0,0.1}
\definecolor{verde}{cmyk}{0.92,0,0.59,0.25}
\definecolor{verdec}{cmyk}{0.92,0,0.59,0.15}
\definecolor{verdes}{cmyk}{0.92,0,0.59,0.4}
\definecolor{Gray}{gray}{0.95}

\font\tenrsfs=rsfs10 at 12pt
\font\sevenrsfs=rsfs7
\font\fiversfs=rsfs5
\newfam\rsfsfam
\textfont\rsfsfam=\tenrsfs
\scriptfont\rsfsfam=\sevenrsfs
\scriptscriptfont\rsfsfam=\fiversfs

\usepackage{chngpage} 

\newcommand{\bpm}{\begin{pmatrix}}      
\newcommand{\epm}{\end{pmatrix}}

\definecolor{lightgray}{rgb}{0.83, 0.83, 0.83}
\definecolor{lightpurp}{rgb}{0.901,0.796,0.882}

\definecolor{colorMag1}{RGB}{255, 255, 178} 
\definecolor{colorMag2}{RGB}{254, 204, 92}  
\definecolor{colorMag3}{RGB}{253, 141, 60}  
\definecolor{colorMag4}{RGB}{240, 59, 32}   
\definecolor{colorMag5}{RGB}{189, 0, 38}    

\newcommand{\bea}{\begin{eqnarray}}
\newcommand{\eea}{\end{eqnarray}}
\newcommand{\beq}{\begin{equation}}
\newcommand{\eeq}{\end{equation}}

\usepackage{tabularx}

\numberwithin{equation}{section}

\definecolor{niceRed}{rgb}{0.7,0.1,0.1}

\usepackage{tikz}
\usetikzlibrary{calc}

\title{\bf 
Shockwaves and Time Delays\\ in Einstein--Maxwell Effective Field Theory
}

\author[a,b]{Christophe Grojean,}
\emailAdd{christophe.grojean@desy.de}

\author[a,c,d]{Minyuan Jiang,}
\emailAdd{minyuan.jiang@nnu.edu.cn}
\author[a]{Pham Ngoc Hoa Vuong}
\emailAdd{hoa.vuong@desy.de}

\affiliation[a]{Deutsches Elektronen-Synchrotron DESY, Notkestr. 85, 22607 Hamburg, Germany}

\affiliation[b]{Institut für Physik, Humboldt-Universität zu Berlin, 12489 Berlin, Germany}
\affiliation[c]{Department of Physics and Institute of Theoretical Physics, Nanjing Normal University,
Nanjing, 210023, China}
\affiliation[d]{Nanjing Key Laboratory of Particle Physics and Astrophysics,
Nanjing, 210023, China}

\abstract{We derive the shockwave metric in four-dimensional Einstein--Maxwell effective field theory (EFT) by performing an ultra-relativistic boost of the charged black hole solution accompanied by a rescaling of its mass and charge, including leading order EFT corrections. In contrast to the neutral (Schwarzschild) case, where higher derivative operators leave the shockwave geometry unchanged, we show that electrically charged shockwaves receive non-trivial EFT corrections. We then compute the time delay experienced by a probe photon traversing the resulting charged shockwave. We find that two EFT contributions, the correction to the shockwave geometry and the backreaction induced by the probe photon, are both essential for obtaining a physical time delay that is invariant under field redefinitions of the metric.
}

\begin{document}
{
\rightline{DESY-25-186}
\rightline{HU-EP-25/42}
}
\maketitle
\flushbottom

\section{Introduction}
Causality is one of the most fundamental principles in physics, and it provides a powerful tool for constraining effective field theories (EFTs). In the pioneering work of \emph{Adams et al.}~\cite{Adams:2006sv}, causality was implemented in two complementary ways. The first uses the analyticity of the $S$-matrix: when combined with unitarity, locality, and Lorentz invariance, analyticity yields positivity bounds that carve out the allowed region of EFT parameter space. The second approach examines how higher derivative operators modify the equations of motion on classical backgrounds and, in particular, how they affect the propagation speed of small perturbations. Requiring that such perturbations propagate no faster than light in vacuum enforces positivity constraints on the associated Wilson coefficients.

When gravity is included, the first method becomes significantly more subtle: the graviton introduces an unavoidable 
$t$-channel pole that obscures straightforward positivity bounds~\cite{Tokuda:2020mlf,Alberte:2020bdz,Alberte:2020jsk,Bellazzini:2019xts,Caron-Huot:2022ugt}, and in curved spacetime the  usual analyticity properties may fail~\cite{deRham:2020zyh,Hollowood:2007kt}. The second method, based on causal propagation on classical backgrounds, therefore offers valuable complementary insight. However, it raises a further question: what backgrounds provide the sharpest and most reliable causality tests? A particularly important development was the work of \emph{Camanho et al.}~\cite{Camanho:2014apa}, who studied the propagation of photons and gravitons in Einstein--Maxwell EFT on gravitational shockwave backgrounds -- geometries obtained as the ultra-relativistic boosts of the Schwarzschild solution~\cite{Aichelburg:1970dh,Dray:1984ha}\footnote{See Refs.~\cite{CarrilloGonzalez:2023cbf,CarrilloGonzalez:2022fwg,Chen:2021bvg} for causality bounds using homogeneous or spherically symmetric backgrounds.}.
The key observable in this setting is the time delay experienced by the probe field as it traverses the shockwave. This delay contains both the familiar Shapiro time delay\footnote{When light propagates near a massive body (such as a star or planet), it suffers a time delay relative to propagation in flat space~\cite{Shapiro:1964uw}.
There are three standard ways to compute this effect: the geodesic approach in GR, the scattering amplitude approach using the eikonal approximation, and the gravitational shockwaves method; see Ref.~\cite{Alexander:2025gdn} for a recent review.} and the additional contributions generated by higher derivative EFT operators. In the framework of Ref.~\cite{Camanho:2014apa}, causality is implemented by demanding that the total time delay remain positive. This criterion is in the literature commonly referred to as \textit{asymptotic causality}. 
However, in presence of gravity, the physical meaning of this causality criterion are still debated in particular in 4D spacetimes.
Indeed, for the effective operators studied in Ref.~\cite{Camanho:2014apa} (e.g. the Gauss--Bonnet term and the leading photon non-minimal coupling to gravity), it was stressed that causality is always preserved as long as the EFT expansion remains under control. In other words, causality breakdown informs about the EFT cutoff scale beyond which UV theories  must be included.
Another implementation of causality, known as \textit{infrared causality}, requires that the EFT contribution to the time delay, if resolvable within the quantum uncertainty, be positive~\cite{Chen:2021bvg}. Using this criterion, it has been observed in Refs.~\cite{deRham:2020zyh,Chen:2023rar}, that for the operators considered in Ref.~\cite{Camanho:2014apa}, causality does not yield any non-trivial bound within the regime of validity of the EFT expansion.

The recent works of \emph{Cremonini et al.}~\cite{Cremonini:2023epg,Cremonini:2024lxn} extended earlier analyses by studying the propagation of a probe photon through charged shockwaves~\cite{Lousto:1988ua} generated by ultra–relativistic boosts of Reissner--Nordstr\"{o}m (RN) black holes. Because such shockwaves carry both gravitational and electromagnetic (EM) fields, the resulting time delay is sensitive to the full set of leading higher derivative operators in Einstein--Maxwell EFT. Interestingly, in this setup, non-trivial bounds can be obtained for pure Maxwell EFT operators in the limit of infinite Planck mass. These results agree with the positivity bounds derived from dispersion relations in the scattering amplitude approach~\cite{Henriksson:2022oeu}. For finite Planck mass, the flat spacetime positivity bounds receive some corrections from gravitational effects as observed by both shockwave and amplitude methods.
The connection between these two approaches still requires further investigation.

One of the motivations highlighted in Ref.~\cite{Cremonini:2023epg} is to derive bounds on the Wilson coefficients of the effective operators that are relevant for the black hole Weak Gravity Conjecture (WGC)~\cite{Arkani-Hamed:2006emk,Kats:2006xp}, which asserts that certain combination of Wilson coefficients needs to be positive so that extremal back holes have a charge-to-mass ratio greater than unity. Under suitable assumptions, the positivity of such combination of Wilson coefficients can be proven from unitarity and causality~\cite{Hamada:2018dde,Bellazzini:2019xts,Henriksson:2022oeu}. The WGC has also been studied from several complementary perspectives within EFT:~ some works study the EFT contributions to the entropy of thermodynamically stable black holes~\cite{Cheung:2018cwt,Cheung:2019cwi}, while others focus on the EFT corrections to black hole quasi-normal modes~\cite{Melville:2024zjq,Boyce:2025fpr,DiRusso:2025qpf}.
In addition, other works investigate the WGC by combining infrared consistency conditions (analyticity, unitarity, causality) with explicit UV to IR matching calculations~\cite{Cheung:2014ega}.

Since higher derivative operators modify the RN solution~\cite{Kats:2006xp}, it is natural to expect that the corresponding charged shockwave geometries should receive EFT corrections as well. In the first attempt to compute the time delay using charged shockwave in Einstein--Maxwell EFT~\cite{Cremonini:2023epg}, the shockwave obtained from boosting the uncorrected RN metric was used, without accounting for these potentially non-vanishing EFT contributions. While it is well established that the purely gravitational shockwave is not corrected by higher derivative operators~\cite{Horowitz:1989bv,Camanho:2014apa,Chen:2023rar}, it remains an open question whether the same holds for electrically charged shockwaves. If the corrections are non-zero, then a consistent computation of the time delay in Einstein--Maxwell EFT must incorporate them.
The aim of this work is to provide precisely such a systematic and self-consistent computation of the time delay for charged shockwaves in Einstein--Maxwell EFT.

Interestingly, we find that such EFT induced corrections to the charged shockwave metric do exist and contribute to the time delay. In addition, there is another EFT induced backreaction effect that must be included but has been largely overlooked in the literature. The interaction between the probe photon and the background electromagnetic field induces a perturbation of the metric; in the presence of higher derivative operators, this perturbation yields non-vanishing contributions to the time delay. We show that only by including both types of backreaction simultaneously does one obtain a result that is invariant under field redefinitions, as required for consistency within EFT.

This paper is organized as follows. In Section~\ref{sec.2-review}, we define the effective operators of the Einstein–Maxwell EFT and discuss operator redundancies together with the corresponding shifts of Wilson coefficients under field redefinitions. We then review the RN black hole solution and its EFT corrections. In Section~\ref{sec.3-shockwaves}, after briefly recalling how to boost the RN solution to obtain a shockwave geometry, we derive for the first time the EFT corrections to the charged shockwave metric. In Section~\ref{sec.4-time-delay}, we study the time delay experienced by a probe photon on this background, highlighting the second backreaction effect and demonstrating how including it restores the field redefinition invariance of the total time delay. We conclude in Section~\ref{sec.5-conc-outl} with a summary of our results and a discussion of future directions.

\section{Review: Black Hole solutions in Einstein--Maxwell EFT}
\label{sec.2-review}
The purpose of this section is to establish the foundations necessary to construct the gravitational shockwave background in Section~\ref{sec.3-shockwaves} and to calculate the time delays in Einstein–Maxwell EFT in Section~\ref{sec.4-time-delay}.
In this section, we briefly review the RN black hole metric and its corrections arising from four-derivative effective operators. We also discuss operator redundancies under the field redefinitions which serve as a useful systematic cross-check when calculating physical quantities such as the mass--charge ratio of extremal black holes and the time delay experienced by probe particles in shockwave background. We closely follow Refs.~\cite{Kats:2006xp, MatthiasBlau} to set up the conventions and notation.

\subsection{Higher derivative interactions and operator redundancy}
Our interest is the low energy effective field theory (EFT) of gravitons and photons. 
We begin with the four dimensional Einstein--Maxwell effective Lagrangian, including terms with up to four derivatives. 
These effective operators may arise from integrating out heavy charged particles within the QFT framework~\cite{Drummond:1979pp,Bastianelli:2008cu,Bastianelli:2012bz,Dunne:2004nc,Ellis:2020ivx,Larue:2023uyv,Bittar:2024xuc}, or from physics at even higher scales associated with quantum gravity, beyond the QFT description. A systematic construction of the EFT with gravity has been developed in Refs.~\cite{Ruhdorfer:2019qmk,Li:2023wdz}, where operators are organized according to their mass dimension and redundancies are removed using field redefinitions and equations of motion. While this approach provides a comprehensive operator basis, in the present work we adopt a complementary organization based on derivative counting, which is more directly adapted to our analysis of shockwave geometries and time delays. We therefore focus on the following Einstein--Maxwell effective action~\cite{Kats:2006xp}\footnote{ The operator $\tilde{\mathcal{O}}_7$ is not included in~\cite{Kats:2006xp} since it does not modify the black hole solution (as seen in Section~\ref{sec.2-review}). We include it here because it will contribute to the time delay.}
\begin{align}
    \label{Einstein-Maxwell-action}
    \mathcal{S}_{\rm eff}
    &= \int d^4x \, \sqrt{-g} \left[ \mathcal{L}_{_{\rm Einstein-Maxwell }} + \Delta \mathcal{L}(c_i\mathcal{O}_i) \right] 
    \nonumber \\
    &= \int d^4x \, \sqrt{-g} \bigg[ \dfrac{R}{2\kappa^2} -\dfrac{1}{4}F_{\mu\nu}F^{\mu\nu} +c_1 R^2+c_2 R_{\mu\nu}R^{\mu\nu} +c_3 R_{\mu\nu\lambda\sigma}R^{\mu\nu\lambda\sigma} 
    \nonumber\\
    & +c_4 R F_{\mu \nu} F^{\mu \nu} +c_5 R^{\mu\nu} F_{\mu \lambda} F_{\nu}^{~\lambda} +c_6 R^{\mu \nu \lambda \sigma} F_{\mu \nu} F_{\lambda\sigma}
    +c_7\big(F_{\mu\nu}F^{\mu\nu}\big)^2 + \tilde{c}_7\big(F_{\mu\nu}\tilde{F}^{\mu\nu}\big)^2
    \nonumber\\
    & +c_8\big(\nabla_\mu F_{\nu \lambda}\big)\big(\nabla^\mu F^{\nu\lambda}\big) +c_9\big(\nabla_\mu F_{\nu \lambda}\big)\big(\nabla^\nu F^{\mu\lambda}\big) \bigg]
    \,,
\end{align}
where $\kappa^2 = 8\pi G_N$, with $G_N$ the Newton constant, and $c_i$ are the Wilson coefficients \footnote{While $\kappa$ has mass dimension $M^{-1}$, the Wilson coefficients have the following mass dimensions: $[c_1,c_2,c_3]=M^0$, $[c_4,c_5,c_6]=M^{-2}$, and $[c_7,\tilde{c}_7,c_8,c_9]=M^{-4}$\,.}. Throughout, we work in the mostly-plus metric signature $(-,+ \cdots,+)$. The convention for the dual field strength tensor is $\tilde{F}^{\mu\nu}=\frac{1}{2} \epsilon^{\mu\nu\alpha\beta}F_{\alpha\beta}$, where $\epsilon^{\mu\nu\alpha\beta} = \sqrt{-g}\, \bar{\epsilon}^{\mu\nu\alpha\beta}$ with $\bar{\epsilon}^{\mu\nu\alpha\beta}$ being the Levi--Civita tensor in flat spacetime. We also adopt the standard conventions of General Relativity, including the assumption of a Levi--Civita connection (metric-compatible and torsion-free). Hence, $\nabla$ denotes the covariant derivative involving only the Christoffel connection. The covariant derivative of a vector field $v_{\nu}$ is defined as follows: 
\begin{align}
    \label{convention: christoffel}
    \nabla_\mu v_\nu = \partial_\mu v_\nu - \Gamma_{\mu\nu}^{\lambda}v_\lambda 
    \,,\quad
    \nabla_\mu v^\nu = \partial_\mu v^\nu + \Gamma_{\mu\lambda}^{\nu} v^\lambda 
    \,.
\end{align}
The Riemann curvature, Ricci tensor, Ricci scalar, and Maxwell field strength tensor are defined as,
\begin{align}
    \label{convetion: riemann}
    \big[ \nabla_\lambda \,, \nabla_\sigma \big]v^\mu = R^{\mu}_{~\,\nu\lambda\sigma} v^{\nu}
    \,,~~
    R_{\mu\nu} = R^{\lambda}_{~\,\mu\lambda\nu}
    \,,~~
    R = g^{\mu\nu}R_{\mu\nu}
    \,;~~
    F_{\mu\nu} = \partial_\mu A^{^{\rm EM}}_\nu - \partial_\nu A^{^{\rm EM}}_\mu
    \,,
\end{align}
with $A^{^{\rm EM}}_{\mu}$ being the electromagnetic gauge potential. 

\paragraph{Power counting.} If one is interested in the region near the event horizon of the extremal RN black hole, $r/\kappa \sim Q\simeq \kappa M\gg 1$, with $\{Q,M\}$ denoting the black hole's charge and mass (see Section~\ref{RN-metric-EFT} for the explicit RN solution), the unperturbed RN solution indicates that each derivative contributes at order $1/Q$. Therefore, all operators in the pure Einstein--Maxwell Lagrangian (the $R$ and $F^2$ terms) involving only two derivatives contribute at order $1/Q^2$, while the remaining EFT operators containing four derivatives contribute at order $1/Q^4$. We will see in Section~\ref{sec.4-time-delay} that higher derivative operators contribute to the time delay with higher power of $1/\rho$, with $\rho$ being the impact parameter of probe particle. For EFT to be valid we generally require $1/\rho \ll \Lambda$ with $\Lambda$ being the cut-off the EFT. The EFT expansion can be viewed as an expansion in $1/(\rho \Lambda)$. This justifies our effective operators as leading modifications to the Einstein--Maxwell theory.

\paragraph{Operator redundancy and field redefinition of the metric.} Before proceeding to the EFT corrections to the RN solution and to the gravitational shockwave solutions obtained from boosted RN black holes, we first review the operator basis given in Eq.~\eqref{Einstein-Maxwell-action}. We note that the use of metric field redefinitions to eliminate redundant operators and simplify the EFT Lagrangian~\eqref{Einstein-Maxwell-action} has been extensively discussed in the literature; see Refs.~\cite{Cheung:2014ega,Cheung:2018cwt,Myers:2009ij}. For completeness, we summarise the key points below.

$\bullet$ First, we consider only the higher derivative terms that are invariant under CP transformations. We omit the operator $(\nabla_{\mu}F^{\mu\nu})(\nabla^{\lambda}F_{\lambda\nu})$ because the term $(\nabla_{\mu}F^{\mu\nu})$ vanishes in the presence of the unperturbed solution to the Maxwell equation (see Section~\ref{RN-metric-EFT} for further details). We also omit the operator $F^{\mu\nu}F_{\nu\lambda}F^{\lambda\sigma}F_{\sigma\mu}$ as it is algebraically related to the operators $(F_{\mu\nu}F^{\mu\nu})^2$ and $(F_{\mu\nu}\tilde{F}^{\mu\nu})^2$. More precisely, in $D=4$ dimensions, one can use the following useful identity~\cite{Cheung:2014ega}:
\begin{align}
    2\big(F_{\mu\nu}F^{\mu\nu}\big)^2 + \big(F_{\mu\nu}\tilde{F}^{\mu\nu}\big)^2 =4F_{\mu\lambda} F^{\lambda \nu} F_{\nu\sigma} F^{\sigma\mu}
    \,.
\end{align}

$\bullet$ Second, the operators $(\nabla_\mu F_{\nu \lambda}\big)\big(\nabla^\mu F^{\nu\lambda})$ and $(\nabla_\mu F_{\nu \lambda}\big)\big(\nabla^\nu F^{\mu\lambda})$ can be removed, and their corresponding Wilson coefficients $c_8, \, c_9$ can also be absorbed into $c_5$ and $c_6$ by making use of the Bianchi identities and integration by parts~\cite{Cremonini:2009sy}\footnote{Notice that Ref.~\cite{Cremonini:2009sy} uses a different sign convention for the Riemann curvature.},
\begin{align}
    \label{eq: IBP-c8-c9}
    &c_8\big(\nabla_\mu F_{\nu \lambda}\big)\big(\nabla^\mu F^{\nu\lambda}\big)  +c_9\big(\nabla_\mu F_{\nu \lambda}\big)\big(\nabla^\nu F^{\mu\lambda}\big) 
    \nonumber \\
    & \hspace{1cm} = (2c_8 + c_9)\left[ -F_{\mu\nu}\nabla^{\nu}\nabla_{\lambda}F^{\mu\lambda} - R^{\mu\nu}F_{\mu\lambda}F_{\nu}^{~\lambda} +\dfrac{1}{2}R^{\mu\nu\lambda\sigma}F_{\mu\nu}F_{\lambda\sigma} \right]
    \,.
\end{align}
Again, the first term on the RHS of Eq.~\eqref{eq: IBP-c8-c9}, $F_{\mu\nu}\nabla^{\nu}\nabla_{\lambda}F^{\mu\lambda}$, can be dropped since $\nabla_{\lambda}F^{\mu\lambda}$ vanishes by itself in the unperturbed solution. The remaining terms can be absorbed into $c_5$ and $c_6$.

$\bullet$ Third, 
we can make use of the fact that the $4D$ Gauss--Bonnet term,
\begin{align}
    \label{eq:gauss-bonnet-term}
    R_{\mu\nu\lambda\sigma}R^{\mu\nu\lambda\sigma} -4R_{\mu\nu}R^{\mu\nu} +R^2
    \,,
\end{align}
is a total derivative, so that we can express the operator $R_{\mu\nu\lambda\sigma}R^{\mu\nu\lambda\sigma}$ in terms of the Ricci tensor and Ricci scalar, and subsequently absorb the coefficient $c_3$ into $c_1$ and $c_2$.

$\bullet$ Last, we can use the following relation from Einstein equation in the pure Einstein--Maxwell Lagrangian of Eq.~\eqref{Einstein-Maxwell-action} to eliminate $\mathcal{O}_{\{1,2,4,5\}}$
\begin{align}
    \label{eq: einstein-eq-tree}
    R &= \big(T^{(0)}\big)^{\mu}_{\mu} = 0\,,\\
    R_{\mu\nu}  &= \big(T^{(0)}\big)_{\mu\nu}
    = \kappa^2 \left( F_{\mu\alpha}F_{\nu}^{~\,\alpha} -\dfrac{1}{4}g_{\mu\nu}F_{\alpha\beta}F^{\alpha\beta}\right)
    \,,\label{eq: einstein-eq-tree2}
\end{align}
where $\big(T^{(0)}\big)_{\mu\nu}$ is the energy--momentum tensor in the pure theory. This leads to 
\begin{align}
\label{eq:implication-EOM-graviton-tree-1}
    \mathcal{O}_1&=R^2 \rightarrow 0\,,\\
    \label{eq:implication-EOM-graviton-tree-2}
\mathcal{O}_2&=R_{\mu\nu}R^{\mu\nu} \rightarrow \dfrac{\kappa^4}{4}(F_{\mu\nu}F^{\mu\nu})^2 + \dfrac{\kappa^4}{4}(F_{\mu\nu}\tilde{F}^{\mu\nu})^2\,,\\
\label{eq:implication-EOM-graviton-tree-4}
\mathcal{O}_4&=RF_{\mu\nu}F^{\mu\nu} \rightarrow 0\,,\\
\label{eq:implication-EOM-graviton-tree-5}
\mathcal{O}_5&=R^{\mu\nu}F_{\mu\lambda}F_\nu^{~\lambda} \rightarrow \dfrac{\kappa^2}{4}(F_{\mu\nu}F^{\mu\nu})^2 + \dfrac{\kappa^2}{4}(F_{\mu\nu}\tilde{F}^{\mu\nu})^2\,.
\end{align}
Such procedure relying on the use of leading-order equations of motion is known to be equivalent to performing a field redefinition of the metric~\cite{Cheung:2018cwt,Hamada:2018dde,Arkani-Hamed:2021ajd,Boyce:2025fpr}. For the later use we write down explicitly the field redefinition that eliminates $\mathcal{O}_5$ and shifts the coefficient of $c_7$ and $\tilde{c}_7$: 
\begin{align}
    \label{eq: field-redefinition-c5}
    g_{\mu\nu}\rightarrow g_{\mu\nu}+\delta g^{(c_5)}_{\mu \nu} 
    \,, ~\,\text{where}~\,
    \delta g^{(c_5)}_{\mu \nu}=2c_5\kappa^2 F_{\mu\lambda}F_{\nu}^{~\,\lambda} - c_5\kappa^2 g_{\mu\nu}F_{\alpha\beta}F^{\alpha\beta}
    \,.
\end{align}
One can check under such field redefinition the pure Einstein--Maxwell action changes 
\begin{align}
    \label{eq: shifted-action-c5}
    \dfrac{1}{\sqrt{-g}} \delta\mathcal{L} 
    &= \delta g_{\mu\nu}^{(c_5)} \dfrac{1}{2\kappa^2} \left[ R^{\mu\nu} -\dfrac{1}{2}g^{\mu\nu}R -\kappa^2\left( F^{\mu\alpha}F^{\nu}_{~\,\alpha} -\dfrac{1}{4}g^{\mu\nu}F_{\alpha\beta}F^{\alpha\beta} \right) \right]\nonumber\\
    &=-c_5\mathcal{O}_5 +\kappa^2\dfrac{c_5}{4}\big(\mathcal{O}_7+\mathcal{\tilde{O}}_7 \big) \,.
\end{align}

Putting everything together (as discussed around Eqs.~\eqref{eq: IBP-c8-c9}, \eqref{eq:gauss-bonnet-term}, \eqref{eq:implication-EOM-graviton-tree-1}--\eqref{eq:implication-EOM-graviton-tree-5}), we are left with three independent operators, which can be chosen as
\begin{align}
    \label{eq: eft-basis}
    \dfrac{1}{\sqrt{-g}}\mathcal{L}
    = \dfrac{R}{2\kappa^2} -\dfrac{1}{4}F_{\mu\nu}F^{\mu\nu}
    +c_6^{\prime} R^{\mu \nu \lambda \sigma} F_{\mu \nu} F_{\lambda\sigma}
    +c_7^{\prime} \big(F_{\mu\nu}F^{\mu\nu}\big)^2 
    +\tilde{c}_7^{\prime} \big(F_{\mu\nu}\tilde{F}^{\mu\nu}\big)^2
    \,,
\end{align}
where the Wilson coefficients shift as follows:
\begin{align}
    \label{eq: results-metric-redef}
    \
    &\big\{  c_1, c_2, c_3, c_4, c_5, c_8, c_9 \big\} \rightarrow \big\{ c'_1, c'_2, c'_3, c'_4, c'_5, c'_8, c'_9 \big\} = 0 \,;
    \nonumber \\
    &c_6 \rightarrow c'_6 =c_6 +\dfrac{2c_8 + c_9}{2} \,; 
    \nonumber \\
    &c_7 \rightarrow c'_7 =c_7+\kappa^4 \left(\dfrac{c_2}{4}+c_3 \right) +\kappa^2 \left(\dfrac{c_5}{4}-\dfrac{2c_8+c_9}{4} \right) \,; 
    \\
    &\tilde{c}_7 \rightarrow \tilde{c}'_7 =\tilde{c}_7+\kappa^4 \left( \dfrac{c_2}{4}+c_3 \right)+\kappa^2 \left( \dfrac{c_5}{4}-\dfrac{2c_8+c_9}{4} \right) \,.
    \nonumber 
\end{align}
Note that the complete metric field redefinitions in general $D$ dimensions are listed in Refs.~\cite{Cheung:2018cwt, Myers:2009ij}. 
The crucial point is that any physical quantity must be independent of the choice of operator basis and, therefore, must remain invariant under reparametrisations of the field variables. For example, the EFT corrections to the mass--charge ratio~\cite{Kats:2006xp}, to the entropy $\Delta S$~\cite{Cheung:2018cwt}, and to the quasi-normal modes~\cite{Boyce:2025fpr} of RN black holes have been shown to be invariant under such metric field redefinitions. A similar argument can apply for the EFT corrections to the time delays.
Even if there are only three physical independent operators with up to four derivatives, in the following, we will work in the non-minimal basis given in Eq.~\eqref{Einstein-Maxwell-action} since our goal is to use the invariance under the reparametrisations listed in Eq.~\eqref{eq: results-metric-redef} as a consistency test of our results and to highlight certain subtleties in the intermediate steps of the calculations.

\subsection{EFT corrections to Reissner--Nordstr\"{o}m black hole metric}
\label{RN-metric-EFT}
In General Relativity, the gravitational collapse of a star with small but nonzero net charge and slight asymmetries can form a final black hole whose external field is fully determined by the mass $M$, charge $Q$, and intrinsic angular momentum $L$ of the progenitor star. Here, we only consider the non-rotating charged black holes, described by the RN solution of the Einstein--Maxwell equations. Following the algorithm of Ref.~\cite{Kats:2006xp}, one can compute the EFT corrections to the RN metric order by order in the Wilson coefficients $c_i$. We summarise below the key steps of the prescription of Ref.~\cite{Kats:2006xp} for deriving the first order corrections in $c_i$ to the RN metric. 

$\bullet$ We begin with the ansatz for the RN solution, adopting the \textit{standard form} of the spherically symmetric metric,
\begin{align}
    \label{met-spherical-form}
    ds^2= -A(r)\, dt^2 + B_{\rm inv}^{-1}(r)\, dr^2+r^2 \big( d\theta^2+\sin^2\theta d\phi^2 \big)
    \,,
\end{align}
where 
\begin{align}
    \label{eq: ansatz-metric}
     A(r) = A^{(0)}(r) + A^{(c_i)}(r)
    \,,\quad
  B_{\rm inv}(r) = B_{\rm inv}^{(0)}(r) + B_{\rm inv}^{(c_i)}(r)
    \,.
\end{align}
The ansatz for the 1-form electromagnetic gauge potential of the RN solution, assuming static spherical symmetry and an electric monopole configuration\footnote{For simplicity, we do not consider magnetic or dyonic black holes in this paper.}, is
\begin{align}
    \label{eq: ansatz-gauge}
    A^{^{\rm EM}} \equiv A^{^{\rm EM}}_{\mu}\, dx^{\mu} = \phi(r)\,dt 
    \,,\quad \text{where} \quad
    \phi(r) = \phi^{(0)}(r) + \phi^{(c_i)}(r)
    \,.
\end{align}
Here, $A^{(0)}(r),~B^{(0)}(r)$ and $\phi^{(0)}(r)$ denote the unperturbed solutions to the coupled Einstein--Maxwell equations, while the perturbative solutions $ A^{(c_i)}(r),~B_{\rm inv}^{(c_i)}(r)$ and $\phi^{(c_i)}(r)$ represent the first order EFT corrections induced by the higher derivative operators.

$\bullet$ Using spherical symmetry, one can express $B_{\rm inv}(r)$ and $A(r)B_{\rm inv}^{-1}(r)$ in terms of the non-vanishing components of the Ricci tensor $R_{\mu\nu}$. We also assume that, in the limit $r \rightarrow \infty$, the asymptotic behaviour of the RN metric is the Schwarzschild (vacuum) solution. From explicit calculation of curvature tensors from the metric Eq.~\eqref{eq: ansatz-metric}, we find
\begin{align}
    \label{eq: metric-solv-Ricci}
    B_{\rm inv}(r)
    &= 1 -\dfrac{\kappa^2 m}{r} -\dfrac{1}{r} \int_r^\infty dr ~ r^2 \left[~ \dfrac{1}{2} (R_t^t-R_r^r)-R_\theta^\theta ~\right]
    \,, \\
    A(r)B_{\rm inv}^{-1}(r) &= \exp \left[~ \int^\infty_r dr ~ r\big( R^t_t-R^r_r \big) B_{\rm inv}^{-1}(r) \right]
    \,, \\
    R_\theta^\theta &= R_\phi^\phi\,,
\end{align}
where $m \equiv M/4\pi$. By making use of Einstein's equation\footnote{We treat EFT operators as a part of $\mathcal{S}_{\rm matter}$\,, i.e. $\mathcal{S}_{\rm matter} = \int d^4x \, \sqrt{-g}\left[ -\dfrac{1}{4}F_{\mu\nu}F^{\mu\nu} + \Delta\mathcal{L}(c_i\mathcal{O}_i) \right] $\,.}, 
\begin{align}
    \label{eq: einstein-eq-generic}
    R_{\mu\nu}=\kappa^2 \left(T_{\mu\nu}-\dfrac{1}{2}T g_{\mu\nu}\right)
    ~\,\text{with}~\, T \equiv T^{\alpha}_{\alpha}
    \,, ~\,\text{and}~\, T_{\mu\nu} = -\dfrac{2}{\sqrt{-g}}\dfrac{\delta \mathcal{S_{\rm matter}}}{\delta g^{\mu\nu}}
    \,,
\end{align}
the RN metric components can be written in terms of the EMT 
\begin{align}
    \label{eq: RN-solve-generic} 
    B_{\rm inv}(r) &= 1-\dfrac{\kappa^2 m}{r}-\dfrac{\kappa^2}{r}\int_r^\infty dr ~ r^2\, T_t^t
    \,, \\
    A(r)B_{\rm inv}^{-1}(r) &= \exp\left[ \kappa^2 \int^\infty_r dr ~ r(T^t_t-T^r_r)B_{\rm inv}^{-1}(r) \right] \,. 
    \label{eq: RN-solve-generic-2} 
\end{align}
To obtain the RN solution explicitly, one first solves the Maxwell equation for $\phi(r)$ and then substitutes into Eqs.~\eqref{eq: RN-solve-generic},~\eqref{eq: RN-solve-generic-2}. This procedure holds in general, and we can apply it iteratively to obtain the higher order EFT corrections to the RN metric.

\paragraph{Unperturbed Reissner--Nordstr\"{o}m metric.} 
The unperturbed RN solution follows from the pure Einstein–Maxwell Lagrangian in Eq.~\eqref{Einstein-Maxwell-action}. 
First, we note that $(T^{(0)})^t_t=(T^{(0)})^r_r$ upon substituting the ansatz~\eqref{eq: ansatz-gauge} into $\big(T^{(0)}\big)_{\mu\nu}$ in Eq.~\eqref{eq: einstein-eq-tree2}. Using Eq.~\eqref{eq: RN-solve-generic-2}, we obtain
\begin{equation}
    A^{(0)}(r)/B_{\rm inv}^{(0)}(r) = 1.
    \label{eq:A/B_0}
\end{equation}
With the ansatz Eqs.~\eqref{met-spherical-form}--\eqref{eq: ansatz-gauge} of the metric and gauge field, the equation of motion (EOM) for the gauge field (the Maxwell equation) becomes
\begin{align}
    \label{eq: EOM-gauge-unperturbed}
    \nabla_\nu F^{\mu\nu} 
    &=\dfrac{1}{A(r)/B_{\rm inv}(r)}\left[ \dfrac{2\phi'(r)}{r}+\phi''(r)\right] -\dfrac{\big[ A(r)/B_{\rm inv} \big]'}{2\big[A(r)/B_{\rm inv}\big]^2}\phi'(r)
    \nonumber \\
    &\approx \dfrac{2\big[\phi^{(0)}(r)\big]'}{r} +\big[\phi^{(0)}(r)\big]''
    =0\,.
\end{align}
An explicit solution for $\phi^{(0)}(r)$ under the electric monopole assumption is
\begin{align}
    \label{eq: unperturbed-phi}
    \phi^{(0)}(r) = \dfrac{q}{r}\; , \;\;\; (T^{(0)})^\mu_\nu={\rm diag}(-1,-1,1,1)\dfrac{q^2}{2r^4}
    \,,
\end{align}
where $q = Q/4\pi$. Using Eqs.~\eqref{eq: RN-solve-generic} and \eqref{eq:A/B_0}, one finds the unperturbed RN solution,
\begin{align}
    A^{(0)}(r)=B_{\rm inv}^{(0)}(r) = 1-\dfrac{\kappa^2 m}{r}+\dfrac{\kappa^2 q^2}{2r^2}
    \,. 
\label{eq:rnB0}
\end{align}

\paragraph{EFT corrections to Reissner--Nordstr\"{o}m metric.}
We treat the EFT operators in \eqref{Einstein-Maxwell-action} as perturbations and require that they affect neither the asymptotic mass nor the asymptotic charge of the black hole. These operators not only modify the EOM of gauge field but also contribute to the EMT that we decompose as $T_{\mu\nu} = T^{^{(0)}}_{\mu\nu} + T^{^{(c_i)}}_{\mu\nu}$.
To derive the first order corrections in $c_i$ to the RN metric, we use the unperturbed solutions to obtain the perturbative solution $\phi^{(c_i)}(r)$ of the modified Maxwell equation, and subsequently determine the corrections $ A^{(c_i)}(r)$ and $B_{\rm inv}^{(c_i)}(r)$ of the metric.

In the presence of EFT operators, the EOM for the gauge field is
\begin{align}
    \label{EOM-full}
    \nabla_{\nu}F^{\mu\nu} &= 4c_4\nabla_{\nu}\big( RF^{\mu\nu} \big) + 2c_5\nabla_{\nu}\big(R^{\mu\lambda}F_{\lambda}^{~\,\nu} - R^{\nu\lambda}F_{\lambda}^{~\,\mu} \big) 
    +4c_6\nabla_{\nu}\big( R^{\alpha\beta\mu\nu}F_{\alpha\beta} \big)
    \nonumber \\
    &+8c_7\nabla_{\nu}\big( F_{\lambda\sigma}F^{\lambda\sigma}F^{\mu\nu} \big)
    +8\tilde{c}_7\nabla_{\nu}\big( \tilde{F}^{\mu\nu} F_{\alpha\beta}\tilde{F}^{\alpha\beta} \big)
    \nonumber \\
    &-4c_8\nabla_{\nu}\square F^{\mu\nu}
    -2c_9\nabla_{\nu}\nabla_{\lambda}\big( \nabla^{\mu}F^{\lambda\nu} -\nabla^{\nu}F^{\lambda\mu} \big)
    \,,
\end{align}
where we denoted $\square = \nabla^2$. The EMT also receives corrections,  
\begin{align}
    \label{EMT-full}
    T_{\mu\nu} &\equiv -\dfrac{2}{\sqrt{-g}}\dfrac{\delta \mathcal{S_{\rm matter}}}{\delta g^{\mu\nu}}
    = T_{\mu\nu}^{^{(0)}} +  T_{\mu\nu}^{^{(c_i)}}
    \nonumber \\
    &= F_{\mu\alpha}F_{\nu}^{~\alpha} -\dfrac{1}{4}g_{\mu\nu}F_{\alpha\beta}F^{\alpha\beta}
    +c_1\big( g_{\mu\nu}R^2 - 4RR_{\mu\nu} + 4\nabla_{\nu}\nabla_{\mu}R -4g_{\mu\nu}\square R \big)
    \nonumber \\
    &+c_2\big( g_{\mu\nu}R_{\alpha\beta}R^{\alpha\beta} +4\nabla_{\alpha}\nabla_{\nu}R^\alpha_\mu -2\square R_{\mu\nu} -g_{\mu\nu}\square R -4R^\alpha_\mu R_{\alpha\nu} \big)
    \nonumber \\
    &+c_3\big( g_{\mu\nu}R_{\alpha\beta\lambda\sigma}R^{\alpha\beta\lambda\sigma} -4R_{\mu\alpha\beta\lambda}R_{\nu}^{~\,\alpha\beta\lambda} -8\square R_{\mu\nu} +4\nabla_\nu \nabla_\mu R +8R^\alpha_\mu R_{\alpha\nu} -8R^{\alpha\beta}R_{\mu\alpha\nu\beta} \big)
    \nonumber \\
    &+c_4\big( g_{\mu\nu}RF_{\alpha\beta}F^{\alpha\beta} -2R_{\mu\nu}F_{\alpha\beta}F^{\alpha\beta} -4RF_{\mu\alpha}F_{\nu}^{~\alpha} +2\nabla_\mu \nabla_\nu F_{\alpha\beta}F^{\alpha\beta} -2g_{\mu\nu}\square F_{\alpha\beta}F^{\alpha\beta} \big)
    \nonumber \\
    &+c_5\big( g_{\mu\nu}R^{\alpha\lambda}F_{\alpha\beta}F_\lambda^{~\,\beta} -4R_{\nu\alpha}F_{\mu\beta}F^{\alpha\beta} -2R^{\alpha\beta}F_{\alpha\mu}F_{\beta\nu} -g_{\mu\nu}\nabla_\alpha \nabla_\beta F^{\alpha\lambda}F^{\beta}_{~\lambda} 
    \nonumber \\ 
    &\quad\quad~~ +2\nabla_\alpha \nabla_\nu F_{\mu\beta}F^{\alpha\beta} -\square F_{\mu\alpha}F_{\nu}^{~\alpha} \big)
    \nonumber \\
    &+c_6\big( g_{\mu\nu}R^{\alpha\beta\lambda\sigma}F_{\alpha\beta}F_{\lambda\sigma} -6R^\alpha_{~\,\mu\beta\lambda}F_{\alpha\nu}F^{\beta\lambda} -4\nabla_\beta \nabla\alpha F^\alpha_{~\mu}F^\beta_{~\nu} \big)
    \nonumber \\
    &+c_7\big( g_{\mu\nu}(F_{\alpha\beta}F^{\alpha\beta})^2 -8F_{\mu\lambda}F_\nu^{~\lambda}F_{\alpha\beta}F^{\alpha\beta} \big) 
    \nonumber \\
    &+\tilde{c}_7\big( g_{\mu\nu}(F_{\alpha\beta}\tilde{F}^{\alpha\beta})^2 + 16F_\mu^{~\alpha}F_{\nu\alpha}F_{\lambda\sigma}F^{\lambda\sigma} - 32F_\mu^{~\alpha}F_\nu^{~\beta}F_{\beta\lambda}F_{\alpha}^{~\lambda} \big)
    \nonumber \\
    &+c_8\big( g_{\mu\nu}(\nabla_\lambda F_{\alpha\beta})(\nabla^\lambda F^{\alpha\beta}) -2(\nabla_\mu F_{\alpha\beta})(\nabla_\nu F^{\alpha\beta}) -4(\nabla_\alpha F_{\beta\mu})(\nabla^\alpha F^\beta_{~\nu}) 
    \nonumber \\
    &\quad\quad~~ + 4\nabla_\alpha F_{\nu\beta} \nabla^\alpha F_\mu^{~\beta} +4\nabla_\alpha F_{\nu\beta} \nabla_\mu F^{\alpha\beta} -4\nabla_\alpha F^\alpha_{~\,\beta} \nabla_\nu F_\mu^{~\beta} \big)
    \nonumber \\
    &+c_9\big( g_{\mu\nu}(\nabla_\alpha F_{\beta\lambda})(\nabla^\beta F^{\alpha\lambda}) -4(\nabla_\mu F^{\alpha\beta})(\nabla_\alpha F_{\nu\beta}) -2(\nabla_\alpha F_{\beta\mu})(\nabla^\beta F^\alpha_{~\,\nu}) 
    \nonumber \\
    &\quad\quad~~ +2\nabla_\alpha F_{\nu\beta} \nabla^\alpha F_\mu^{~\beta} +2\nabla_\alpha F_{\nu\beta} \nabla_\mu F^{\alpha\beta} -2\nabla_\alpha F^\alpha_{~\,\beta} \nabla_\nu F_\mu^{~\beta}
    \big)
    \,.
\end{align}
Since the EFT-induced corrections $ T_{\mu\nu}^{^{(c_i)}}$ contribute to the RN solution Eq.~\eqref{eq: RN-solve-generic} and Eq.~\eqref{eq: RN-solve-generic-2}, the left-hand side of modified Maxwell equation Eq.~\eqref{EOM-full} must be expanded to linear order in $c_i$. 
Using the unperturbed solutions $\big\{A^{(0)}(r),~B_{\rm inv}^{(0)}(r),~\phi^{(0)}(r)\big\}$, we then obtain the perturbative solution $\phi^{(c_i)}(r)$ of the modified Maxwell equation,
\begin{align}
    \phi^{(c_i)}(r) &= \dfrac{q}{5r^5}\Bigg[ ~c_2\big(\kappa^4q^2 \big) 
    +c_3\big(4\kappa^4q^2 \big) + c_4\big(10\kappa^2q^2\big) + c_5\big(\kappa^2q^2\big) + c_6\big(-9\kappa^2q^2 + 10\kappa^2mr \big)
    \nonumber \\
    &-c_7\big(16 q^2\big) + c_8\big(-11\kappa^2q^2 + 10\kappa^2mr \big) + c_9\left(-\dfrac{11}{2}\kappa^2q^2 + 5\kappa^2mr \right)  \Bigg]  
    \,.
\end{align}

One can directly obtain the perturbative solution $B_{\rm inv}^{(c_i)}(r)$ from Eq.~\eqref{eq: RN-solve-generic}, where the two contributions to the EMT, $T_{\mu\nu}^{^{(0)}} $ and $  T_{\mu\nu}^{^{(c_i)}}$ are evaluated using $\phi^{(c_i)}(r)$ and $\phi^{(0)}(r)$, respectively. The full RN solution including EFT corrections then reads, 
\begin{align}
  \label{RNmetric-B-full}
  B_{\rm inv}^{(c_i)}(r) &= 
  -\dfrac{q^2}{r^6} \Bigg[
  \, c_2\dfrac{\kappa^4}{5}\big( 20r^2 + 6\kappa^2 q^2 - 15\kappa^2 m r  \big)
  \nonumber \\
  &+c_3\dfrac{4\kappa^4}{5}\big( 20 r^2 + 6\kappa^2 q^2 - 15\kappa^2 m r   \big)
  +c_4 \kappa^2\big(16r^2 + 6\kappa^2 q^2 - 14\kappa^2 m r  \big)
  \nonumber \\
  &+c_5\dfrac{\kappa^2}{5}\big( 30r^2 + 11\kappa^2 q^2 - 25\kappa^2 m r  \big)
  +c_6\dfrac{\kappa^2}{5}\big( 40 r^2 + 16\kappa^2 q^2 - 35\kappa^2 m r  \big) +c_7\dfrac{4\kappa^2}{5}q^2 
  \nonumber \\
  &+c_8\dfrac{\kappa^2}{5}\big( -20r^2 - 6\kappa^2 q^2 + 15\kappa^2 m r  \big)
  +c_9\dfrac{\kappa^2 }{10}\big( -20 r^2 - 6\kappa^2 q^2  +15\kappa^2 m r \big) \Bigg]
    \,,
\end{align}
and
\begin{align}
    \label{RNmetric-A-full}
  A^{(c_i)}(r) &= 
  -\dfrac{q^2}{r^6} \Bigg[ 
  \, c_2\dfrac{\kappa^4}{5}\big( 10r^2 +\kappa^2 q^2 - 5\kappa^2 m r  \big)
  \nonumber \\
  &+c_3\dfrac{4\kappa^4}{5}\big( 10r^2 +\kappa^2 q^2 - 5\kappa^2 m r   \big)
  +c_4 \kappa^2\big( -4r^2 -4\kappa^2 q^2 +6\kappa^2 m r  \big)
  \nonumber \\
  &+c_5\dfrac{\kappa^2}{5}\big( -4\kappa^2 q^2 +5\kappa^2 m r  \big)
  +c_6\dfrac{\kappa^2 }{5}\big( 10r^2 +\kappa^2 q^2 - 5\kappa^2 m r  \big)
  +c_7\dfrac{4\kappa^2 }{5}q^2 
  \nonumber \\
  &+c_8\dfrac{\kappa^2 }{5}\big( 10 r^2 +9\kappa^2 q^2 - 15\kappa^2 m r  \big)
  +c_9\dfrac{\kappa^2 }{10}\big( 10 r^2 +9\kappa^2 q^2 - 15\kappa^2 m r \big) \Bigg]
    \,.
\end{align}
These results were first derived in Ref.~\cite{Kats:2006xp}. We observe that all EFT corrections to the metric are proportional to the black hole charge, implying that neutral (Schwarzschild) black holes are insensitive to the operators considered here~\footnote{This can be understood  from the fact that there is no pure gravitational operators at four derivative level in 4D. The Schwarzschild metric can receive corrections from operators involving higher derivatives, such as $(R_{\mu\nu\rho\sigma}R^{\mu\nu\rho\sigma})^2,~(R_{\mu\nu\rho\sigma}\tilde{R}^{\mu\nu\rho\sigma})^2$~\cite{Arkani-Hamed:2021ajd}.}. It can be verified that near the horizon of extremal black hole these corrections are suppressed by factor of $1/Q^2$ compared to the unperturbed solution. We also note that the operator $\tilde{\mathcal{O}}_7$ modifies neither the gauge field nor the metric. 
This follows from the fact that, under the gauge field ansatz in~\eqref{eq: ansatz-gauge}, only the $F_{tr}$ (electric) components of the field strength tensor are non-vanishing, thus $F_{\mu\nu} \tilde{F}^{\mu\nu}$ identically vanishes in this configuration.

\paragraph{Mass--Charge ratio of extremal RN black holes.} From the unperturbed RN solution~\eqref{eq:rnB0}, one finds that extremal black holes feature a degenerate horizon located at radius $r_{_H}=\kappa^2 m/2$ at which the RN metric exhibits a coordinate singularity, and they satisfy the extremal condition $\kappa m/\sqrt{2}q =1$. 
To obtain the EFT corrections to the mass--charge ratio of extreme black holes, we solve $A(r)=0$ given in Eq.~\eqref{RNmetric-A-full} perturbatively and requiring all horizons remain degenerate, we then obtain~\cite{Kats:2006xp},
\begin{align}
    \label{RN: mass-charge-ratio}
    \dfrac{\kappa}{\sqrt{2}}\dfrac{m}{q}=1-\dfrac{2}{5q^2}\left( 2c_2+8c_3+\dfrac{2c_5}{\kappa^2}+\dfrac{2c_6}{\kappa^2}+\dfrac{8c_7}{\kappa^4}-\dfrac{2c_8}{\kappa^2}-\dfrac{c_9}{\kappa^2} \right)
    \,.
\end{align}
We can explicitly verify that the mass--charge relation given in Eq.~\eqref{RN: mass-charge-ratio} is invariant under field redefinitions and under the Wilson coefficient shifts in Eq.~\eqref{eq: results-metric-redef}.
It is worth noting that the weak gravity conjecture~\cite{Arkani-Hamed:2006emk} implies the positivity of the specific combination of Wilson coefficients $c_i$ appearing inside the parentheses of Eq.~\eqref{RN: mass-charge-ratio}.

\section{Gravitational shockwaves in Einstein--Maxwell EFT}
\label{sec.3-shockwaves}
The shockwave metric can be obtained from ultra-relativistic boost of the black hole solution. In this section, we first review the derivation of shockwave from the boost of RN black hole, following Refs.~\cite{Lousto:1988ua,Cremonini:2023epg}. Then we derive corrections due to the effective operators, based on the results in Section~\ref{sec.2-review}.

\subsection{Shockwaves from boosted RN black hole in GR}
We first consider the Lorentz boost to the RN solution. The boost is most easily performed in the isotropic coordinate, where the metric takes the form
\begin{align}
    \label{eq: ansatz-isometric}
    ds^2&=-\bar{A}(\bar{r}) \, dt^2 +\bar{B}(\bar{r}) \, \bigg[ d\bar{r}^2+\bar{r}^2 \big( d\theta^2+\sin^2\theta d\phi^2 \big) \bigg].
\end{align}
Compared with Eq.~\eqref{met-spherical-form}, we find the coordinate transformation $r\rightarrow \bar{r}$ satisfies
\begin{align}
    \left(\dfrac{dr}{d\bar{r}} \right)^2 =B_{\rm inv}(r)\left(\dfrac{r}{\bar{r}}\right)^2
    \,,
    \label{eq:isotrans}
\end{align}
and
\begin{equation}
   \bar{A}(\bar{r})=A\big[ r(\bar{r}) \big]\; , \; \; \bar{B}\big( \bar{r})=(r/\bar{r} \big)^2.
\end{equation}
With the unperturbed RN solution Eq.~\eqref{eq:rnB0}, it is not hard to find the solution
\begin{align}
    r^{(0)} &=\bar{r}+\frac{\kappa^2 m}{2}+\frac{\kappa^4 m^2-2\kappa^2 q^2}{16\bar{r}} \,,\\
    \bar{B}^{(0)}(\bar{r}) &=
    \left( 1+\frac{\kappa^2 m}{2\bar{r}}+\dfrac{\kappa^4 m^2-2q^2\kappa^2}{16\bar{r}^2} \right)^2 \,, \\
    \bar{A}^{(0)}(\bar{r}) &= 1-\dfrac{\kappa^2 m}{r^{(0)}}+\dfrac{\kappa^2 q^2}{2\big[ r^{(0)} \big]^2}  \,.
\end{align}
Now it is natural to define the Cartesian coordinates where $\bar{r}^2=x^2+y^2+z^2$ and the Lorentz boost
\begin{align}
    t'&=\gamma(t+\beta x),  \nonumber\\
    x'&=\gamma(x+\beta t),\nonumber\\ y'&=y,\;\;z'=z,
\end{align}
with $\gamma=1/\sqrt{1-\beta^2}$ being the Lorentz factor. Under the boost, the metric in Eq.~\eqref{eq: ansatz-isometric} becomes 
\begin{align}
    ds^2
    &= \gamma^2(\bar{B}-\bar{A})(dt'-\beta dx')^2+\bar{B}(-dt'^2+dx'^2+dy'^2+dz'^2) 
    \,.
\end{align}
Under infinite boost ($\beta \rightarrow 1, \gamma \rightarrow \infty$), the first term in the metric above becomes divergent. 

$\bullet$ \textbf{Scaling limit:} Following Ref.~\cite{Lousto:1988ua}, in order to render a finite shockwave metric, we rescale the mass and charge of the black hole with the boost parameter $\gamma$, such that
\begin{align}
    \label{scalling-limit}
    m_0 = m\gamma
    \,, ~\,\text{and}~\,
    q_0^2 = q^2\gamma
\end{align}
are constants. With such rescaling, we have
\begin{align}
    &\lim_{\gamma \rightarrow \infty}\bar{B}^{(0)} =1 
    \,,
    \nonumber\\
    &\lim_{\gamma \rightarrow \infty}\gamma^2(\bar{B}^{(0)}-\bar{A}^{(0)}) 
    =\lim_{\gamma \rightarrow \infty}\gamma \left(\frac{2m_0\kappa^2}{\bar{r}}-\frac{3\kappa^2q_0^2}{4\bar{r}^2} \right)
    =  h^{(0)}(\rho) \delta(u) + 2\kappa^2 m_0 \dfrac{1}{|u|}
    \,,
\end{align}
with 
\begin{equation}
h^{(0)}(\rho) \equiv 
-4\kappa^2 m_0\log\rho -\frac{3\pi\kappa^2q_0^2}{4\rho}\,.
\end{equation}
In the last step, we have defined the impact parameter as the transverse radius 
\begin{align}
    \rho\equiv\sqrt{y'^2+z'^2}
\end{align}
and the light-cone coordinates 
\begin{align}
    u\equiv t'-x' 
    \,, ~\,
    v\equiv t'+x'
\end{align}
so that $\lim_{\gamma \rightarrow \infty}\bar{r}=\sqrt{\gamma^2u^2+\rho^2}$, and used the following formula with $\epsilon=(2\gamma^2)^{-1}$~\cite{Cremonini:2023epg}
\begin{align}
\lim_{\epsilon \rightarrow 0}\frac{1}{\sqrt{u^2+2\epsilon\rho^2}}&=-\log\rho^2\delta(u)+\frac{1}{|u|},\\
    \lim_{\epsilon \rightarrow 0}\frac{\epsilon^{n-1/2}}{(u^2+2\epsilon\rho^2)^n}&=\frac{\Gamma(n-1/2)}{\Gamma(n)}\frac{\pi^{1/2}}{(\sqrt{2}\rho)^{2n-1}}\delta(u) \,.
\end{align}
We thus obtain the shockwave metric in Einstein--Maxwell theory
\begin{equation}
 \label{eq:sw-EM}
 ds^{2}
 =
 -dudv+\left(h^{(0)}(\rho) \delta(u) +\frac{2\kappa^2m_0}{|u|}\right) du^2+dy'^2+dz'^2 \,.
\end{equation}
We see that the black hole originally centered at $x=0$ is boosted to a shockwave localized at $u=0$, with its gravitational field supported entirely on the transverse plane. Note that the term $2\kappa^2 m_0/|u|$ does not contribute to the spacetime curvature  and therefore has no physical effect.

Similarly, we can obtain the EM field in the shockwave after the Lorentz boost:
\begin{align}
    \label{eq: EM-shockwave}
    \big(A^{^{\rm EM}} \big)^{(0)} 
    &= \dfrac{q}{r}dt = \dfrac{\gamma q}{r(\bar r)}\big(dt'-\beta dx' \big) \nonumber\\
    &
\overset{\gamma \, \rightarrow \, \infty}{=} 
\dfrac{\sqrt{\gamma}\,q_0}{\sqrt{\gamma^2u^2+\rho^2}}du=\dfrac{(2\epsilon)^{1/4}\, q_0}{\sqrt{u^2+2\epsilon \rho^2}}du
    \,.
\end{align}
We find that under infinite boost the leading component is  $A^{^{\rm EM}}_u=H(u,\rho)\equiv\frac{(2\epsilon)^{1/4}\, q_0}{\sqrt{u^2+2\epsilon \rho^2}}$. Although $H(u,\rho)\rightarrow 0$ as $\epsilon \rightarrow 0$, it still can make a contribution to the time delay from the $\mathcal{O}_7$ and $\tilde{\mathcal{O}}_7$ operators, as discussed in Section~\ref{sec.4-time-delay}.

\subsection{Shockwaves from boosted RN black hole in Einstein--Maxwell EFT}
Now we consider the correction to the shockwave metric due to the effective operators. We start by solving Eq.~\eqref{eq:isotrans}, taking into account the full $B(r)$ function (Eq.~\eqref{RNmetric-B-full}) in the EFT.

Since we are interested in the leading order effects of the Wilson coefficients, we solve this equation perturbatively in $c_i$:
\begin{align}
    r(\bar{r})&=r^{(0)}(\bar{r})+ r^{(c_i)}(\bar{r}) +... \;;\nonumber\\
    B_{\rm inv}(r)&=B_{\rm inv}^{(0)}(r)+B_{\rm inv}^{(c_i)}(r)+...\;,\nonumber\\&=B_{\rm inv}^{(0)}(r^{(0)})+\frac{d}{dr}\left(B_{\rm inv}^{(0)}(r)\right)_{r=r^{(0)}} r^{(c_i)}+B_{\rm inv}^{(c_i)}(r^{(0)})+...\;,
\end{align}
where in $...$ we ignore corrections of $\mathcal{O}(c_i^2)$. To the first order in $c_i$, Eq.~\eqref{eq:isotrans} reads
\begin{equation}
    2\frac{dr^{(0)}}{d\bar{r}}\frac{d r^{(c_i)}}{d\bar{r}}=B_{\rm inv}^{(0)}(r^{(0)})\frac{2r^{(0)}r^{(c_i)}}{\bar{r}^2}+\left(\frac{d}{dr}\left(B_{\rm inv}^{(0)}(r)\right)_{r=r^{(0)}} r^{(c_i)}+B_{\rm inv}^{(c_i)}(r^{(0)})\right)\left(\frac{r^{(0)}}{\bar{r}}\right)^2.\nonumber
\end{equation}
Since in the ultra-relativistic boost $\gamma \rightarrow \infty$, $m$ and $q^2$ are infinitesimal with the rescaling $m=m_0/\gamma$ and $q^2=q_0^2/\gamma$, we can keep only the terms $\sim q^2$ in $B_{\rm inv}^{(c_i)}$ in this equation. The solution is found to be 
\begin{align}
    & r^{(c_i)}
    =-(2 c_2 \kappa^2+6 c_3 \kappa^2+8c_4+3 c_5+4c_6 +2c_8 +c_9)\left(\frac{ -\kappa^4m^2+2\kappa^2q^2+16 \bar{r}^2}{2q^2 \bar{r}}\right) \times \nonumber\\
    &\left[\frac{\kappa^2 m^2+4 m \bar{r}-2q^2}{\kappa^4 m^2+8 \kappa^2 m \bar{r}-2\kappa^2q^2+16 \bar{r}^2}+\frac{-2 \kappa^2 m^2+8 m \bar{r}+2 q^2}{-\kappa^4m^2+2\kappa^2 q^2+16 \bar{r}^2}-\frac{3\sqrt{2}m}{2\kappa q} \log\left(\frac{4\bar{r}+m\kappa^2+\sqrt{2}\kappa q}{4\bar{r}+m\kappa^2-\sqrt{2}\kappa q}\right)\right]
\end{align}
Here the integral constant is chosen such that $\lim_{\bar{r}\rightarrow\infty} \big( r/\bar{r} \big) = 1$. 
Notice that the argument of logarithm is positive from $r^{(0)}=(4\bar{r}+m\kappa^2+\sqrt{2}\kappa q)(4\bar{r}+m\kappa^2-\sqrt{2}\kappa q)/(16\bar{r})>0$.

Having in mind that in the end we will take the ultra-relativistic boost and rescale $(m,q^2)$ to be infinitesimal, we can further simplify above result by expanding over $(m,q^2)$ and keeping only the leading order terms,  which gives
\begin{equation}
    r^{(c_i)}=\frac{\kappa^2 q^2}{4\bar{r}^3}(2 c_2 \kappa^2+6 c_3 \kappa^2+8c_4+3 c_5+4c_6 +2c_8 +c_9).
\end{equation}
With this EFT modified coordinate transformation, we can proceed to calculate 
$\bar{B}=(r/\bar{r})^2$ and take the ultra-relativistic boost, following the same routine as in the previous section while keeping first order contribution of the effective operators. 

In the end, we obtain the following shockwave metric in the EFT:
\begin{align}
    ds^2&=-dudv+\left[\left(h^{(0)}(\rho) +  h^{(c_i)}(\rho) \right)\delta(u)+\frac{2\kappa^2m_0}{|u|}\right]du^2 +dy'^2+dz'^2
    \,,
    \label{eq:sw-EFT}
\end{align}
where 
\begin{align}
     h^{(c_i)}(\rho) 
    &= \dfrac{\pi\kappa^2q_0^2}{4\rho^3} \big( 6c_2 \kappa^2+24 c_3 \kappa^2+3 c_5+8c_6 +2c_8 +c_9 \big)
    \,.
\end{align}
We notice that $ h^{(c_i)}(\rho)\sim q_0^2$, which means that the EFT corrections to the shockwave metric are purely from the EM fields: the effective operators generate extra EMT from the background EM field, which modifies the metric. 

Since the gauge field $\big(A^{^{\rm EM}} \big)^{(c_i)}$ is higher order in $(q^2,m)$ compared to $\big(A^{^{\rm EM}} \big)^{(0)}$, we can safely neglect it under ultra-relativistic boost, and take the gauge field in the shockwave background to be $A^{^{\rm EM}}_u=H(u,\rho)$, as given in Eq.~\eqref{eq: EM-shockwave}.

\section{Time delays in Einstein--Maxwell EFT}
\label{sec.4-time-delay}
\subsection{Time delays from gravitational shockwaves}
It is well known that test particles passing through the gravitational shockwave experience time delays. For particles minimally coupled to gravity, this so-called Shapiro time delay is a purely geometric effect and is therefore universal, i.e. independent of the particle's spin.  
\begin{figure}[h!]
\centering
    \begin{subfigure}{0.5\textwidth}
    \centering
    \includegraphics[width=\linewidth]
    {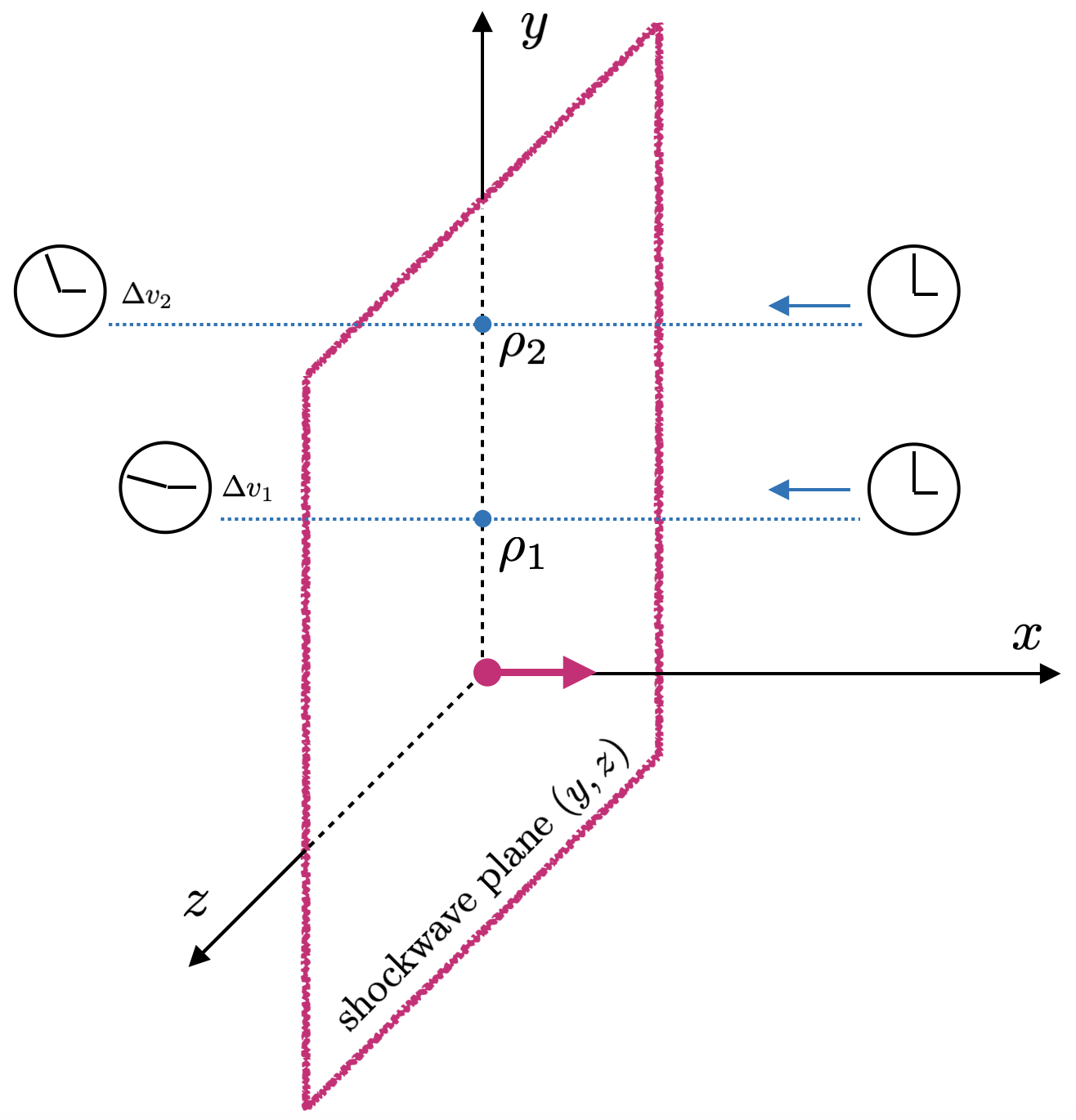}
    \caption{Time delays in Cartesian coordinates}
    \label{fig: time-delay-cartesian}
    \end{subfigure}%
    \begin{subfigure}{0.5\textwidth}
    \centering
    \hfill
    \includegraphics[width=\linewidth]
    {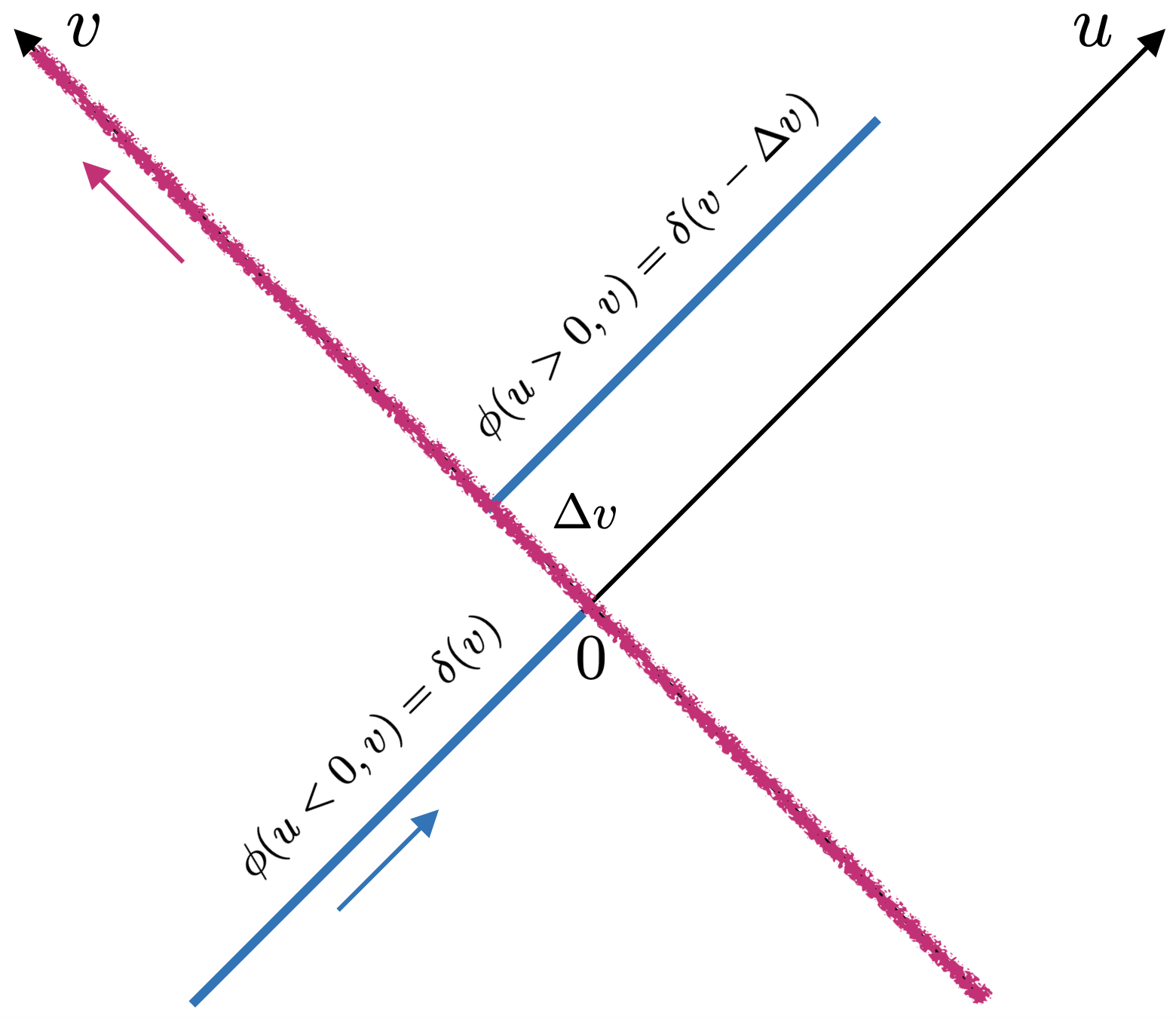}
    \caption{Time delays in light-cone coordinates}
    \label{fig: time-delay-lightcone}
    \end{subfigure}
\caption{\eqref{fig: time-delay-cartesian} Spacetime diagram, in Cartesian coordinates, describing the effect of the shockwave in the $(y,z)$ plane on a probed particle moving backward in the $x$-direction. Two initially synchronised clocks separated by different impact parameters ($y$-direction) become out of sync after crossing the shockwave plane. The clock with smaller impact parameter will experience a larger absolute time delay, i.e. $|\Delta v_1 | > |\Delta v_2|$.
\eqref{fig: time-delay-lightcone} Spacetime diagram, in light-cone coordinates, describing the effect of the shockwave localised at $u = 0$ on a probed particle moving forward in the $u$-direction.
}
\label{fig: time-delay}
\end{figure}

For now, we consider a probe particle described by scalar field $\phi$ propagating in the $x$-direction, see Fig.~\ref{fig: time-delay-cartesian}. In the unperturbed gravitational shockwave background Eq.~\eqref{eq:sw-EM}, the EOM, $\Box \phi=0$,  takes the form
\begin{equation}
   \partial_u \partial_v \phi+h^{(0)}(\rho)\delta(u)\partial^2_v\phi+\frac{2\kappa^2m_0}{|u|} \partial^2_v\phi-\frac{1}{4}(\partial^2_y+\partial^2_z)\phi=0 \,.
   \label{eq:scalar}
\end{equation}
As the probe crosses the shockwave at $u=0$, variation of the wave function in the $u$ direction is much more rapid than in the transverse $(y,z)$ plane, allowing us to neglect the last term. The third term is independent of the impact parameter $\rho$ and produces the same effect for all probe trajectories; this contribution is unphysical and can be removed by a coordinate transformation~\cite{Dray:1984ha}. Keeping only the relevant first two terms, we find that the wave function after crossing the shockwave can be described by the following ansatz
\begin{align}
    \phi(u=0^+, v) 
    = \exp\left[ -\int_{0-}^{0+} du \, h^{(0)}(\rho)\delta(u)\partial_v \right] \phi(u=0^-,v)
    \,,
\end{align} 
which directly solves Eq.~(\ref{eq:scalar}).
Using the delta function to perform the integral over $u$, we obtain
\begin{equation}
    \phi(u=0^+, v)=e^{-h^{(0)}(\rho)\partial_v }\phi(u=0^-,v)=\phi(u=0^-,v- h^{(0)}(\rho)) \,.
\end{equation}
Thus, the wavefunction after crossing the shockwave is equal to the incoming wavefunction evaluated at a shifted value of $v$. This discontinuous shift in $v$ 
\begin{equation}
    \Delta v^{(0)}=h^{(0)}(\rho)=-\left(4\kappa^2m_0\log \rho+\frac{3\pi\kappa^2}{4}\frac{q_0^2}{\rho}\right),
    \label{eq:deltaV_0}
\end{equation}
is the time delay experienced by the probe.
Only differences in time delay between particles with different impact parameters are physically meaningful, so the appearance of the logarithmic term in \eqref{eq:deltaV_0} does not lead to any pathological divergence. Similarly, one can verify that probes of any spin such as photons or gravitons also exhibit this universal geometric contribution to the time delay~\cite{Camanho:2014apa,Cremonini:2023epg}. However, as shown in the following subsection, this universality will not persist in the presence of effective operators: higher derivative interactions can violate the strong equivalence principle, leading to species and polarization dependent time delays.

\subsection{Time delays with 
backreaction effects included}
In this work we study the time delay experienced by a probe photon in Einstein--Maxwell EFT. Throughout the analysis, we discard the redundant operators $\mathcal{O}_{1,2,3,4,8,9}$, whose effects can always be restored by imposing invariance of the final result under field redefinitions. In addition to the minimal basis $\{\mathcal{O}_6, \mathcal{O}_7, \tilde{\mathcal{O}}_7\}$, we will also keep the operator $\mathcal{O}_5$, so that field-redefinition invariance can be cross-checked directly: the different contributions to the photon time delay will not be invariant under the field redefinition but the sum will be, giving some confidence on the validity of the whole computation.

\begin{figure}[t]
\centering
\begin{subfigure}{0.45\textwidth}
    \centering
    \begin{tikzpicture}
    [cross/.style={path picture={ 
  \draw[black]
(path picture bounding box.south east) -- (path picture bounding box.north west) (path picture bounding box.south west) -- (path picture bounding box.north east);
}}]
    \node [draw,thick,circle,cross,minimum width=0.25 cm](B) at (0,-1){}; 
    \draw[thick,graviton] (0,-1) to  [out=90,in=-90] (0,0.6)  ;
    \draw[photon] (0,0.6) to  [out=135,in=-45] (-2,1.5);
    \draw[photon] (0,0.6) to  [out=45,in=-135] (2,1.5) ;
    \node[] at (1.6,0.8) {$f_{\mu\nu}$};
    \node[] at (-1.6,0.8) {$f_{\mu\nu}$};
    \node[] at (0.8,0) {$\big(g^{\rm B}_{\mu\nu}\big)^{(0)}$};
\node[rectangle,fill]  (amp1) at (0,0.65) {};
    \end{tikzpicture}
    \caption{} 
\label{Fig:feynmann_diagram_a}
\end{subfigure}
\begin{subfigure}{0.45\textwidth}
    \centering
    \begin{tikzpicture}
    [cross/.style={path picture={ 
  \draw[black]
(path picture bounding box.south east) -- (path picture bounding box.north west) (path picture bounding box.south west) -- (path picture bounding box.north east);
}}]
 \node [draw,thick,circle,cross,minimum width=0.25 cm](B) at (1.6,-1.5){}; 
 \node [draw,thick,circle,cross,minimum width=0.25 cm](B) at (-1.6,-1.5){}; 
    \draw[photon] (0,-0.2) to  [out=135,in=-45] (-1.8,1.2);
    \draw[thick,photon] (-0,-0.2) to  [out=-135,in=45] (-1.6,-1.5);
    \draw[photon] (0,-0.2) to  [out=45,in=-135] (1.8,1.2)  ;
    \draw[thick,photon] (0,-0.2) to  [out=-45,in=135] (1.6,-1.5);
    \node[] at (1.6,0.5) {$f_{\mu\nu}$};
    \node[] at (-1.6,0.5) {$f_{\mu\nu}$};
    \node[] at (-1.2,-0.5){$F^{\rm B}_{\mu\nu}$};
    \node[] at (1.2,-0.5){$F^{\rm B}_{\mu\nu}$};
    \node[rectangle,fill]  (amp1) at (0,-0.2) {};
    \end{tikzpicture}
    \caption{} \label{Fig:feynmann_diagram_b}
\end{subfigure}
\begin{subfigure}{0.45\textwidth}
    \centering
    \begin{tikzpicture}
     [cross/.style={path picture={ 
  \draw[black]
(path picture bounding box.south east) -- (path picture bounding box.north west) (path picture bounding box.south west) -- (path picture bounding box.north east);
}}]
 \node [draw,thick,circle,cross,minimum width=0.25 cm](B) at (1.6,-1.5){};
  \node [draw,thick,circle,cross,minimum width=0.25 cm](B) at (-1.6,-1.5){};
    \draw[thick,photon] (0,-0.6) to  [out=-135,in=45] (-1.6,-1.5);
    \draw[thick,photon] (0,-0.6) to  [out=-45,in=135] (1.6,-1.5);
    \draw[thick,graviton] (0,-0.6) to  [out=90,in=-90] (0,0.6)  ;
    \draw[photon] (0,0.6) to  [out=135,in=-45] (-2,1.5);
    \draw[photon] (0,0.6) to  [out=45,in=-135] (2,1.5) ;
    \node[] at (-1.2,-0.7) {$F^{\rm B}_{\mu\nu}$};
     \node[] at (1.2,-0.7) {$F^{\rm B}_{\mu\nu}$};
    \node[] at (1.6,0.8) {$f_{\mu\nu}$};
     \node[] at (-1.6,0.8) {$f_{\mu\nu}$};
    \node[] at (0.85,0) {$\big(g^{\rm B}_{\mu\nu}\big)^{(c_i)}$};
\node[rectangle,fill]  (amp1) at (0,-0.65) {};
    \end{tikzpicture}
    \caption{} 
\label{Fig:feynmann_diagram_c}
\end{subfigure}

\begin{subfigure}{0.45\textwidth}
    \centering
    \begin{tikzpicture}
    [cross/.style={path picture={ 
  \draw[black]
(path picture bounding box.south east) -- (path picture bounding box.north west) (path picture bounding box.south west) -- (path picture bounding box.north east);
}}]
 \node [draw,thick,circle,cross,minimum width=0.25 cm](B) at (-1.6,-1){};
  \node [draw,thick,circle,cross,minimum width=0.25 cm](B) at (1.6,-1){};
    \draw[photon] (-1,0) to  [out=135,in=-45] (-2,1);
    \draw[thick,photon] (-1,-0) to  [out=-135,in=45] (-1.6,-1);
    \draw[photon] (1,0) to  [out=45,in=-135] (2,1)  ;
    \draw[thick,photon] (1,0) to  [out=-45,in=135] (1.6,-1);
    \draw[graviton] (-1,0) to [out=0,in=-180] (1,0); 
    \node[] at (0,0.4) {$h_{\mu\nu}$};
    \node[] at (2.,0.5) {$f_{\mu\nu}$};
     \node[] at (-2.,0.5) {$f_{\mu\nu}$};
    \node[] at (-1.8,-0.4){$F^{\rm B}_{\mu\nu}$};
    \node[] at (1.8,-0.4){$F^{\rm B}_{\mu\nu}$};
    \node[rectangle,fill]  (amp1) at (-1,0) {};
    \end{tikzpicture}
    \caption{} 
    \label{Fig:feynmann_diagram_d}
\end{subfigure}
\begin{subfigure}{0.45\textwidth}
    \centering
    \begin{tikzpicture}
    [cross/.style={path picture={ 
  \draw[black]
(path picture bounding box.south east) -- (path picture bounding box.north west) (path picture bounding box.south west) -- (path picture bounding box.north east);
}}]
 \node [draw,thick,circle,cross, minimum width=0.25 cm](B) at (1.6,-1.){};
  \node [draw,thick,circle,cross,minimum width=0.05 cm](B) at (-1.6,-1.){};
    \draw[photon] (-1,0) to  [out=135,in=-45] (-2,1);
    \draw[thick,photon] (-1,-0) to  [out=-135,in=45] (-1.6,-1);
    \draw[photon] (1,0) to  [out=45,in=-135] (2,1)  ;
    \draw[thick,photon] (1,0) to  [out=-45,in=135] (1.6,-1);
    \draw[graviton] (-1,0) to [out=0,in=-180] (1,0); 
   \node[] at (0,0.4) {$h_{\mu\nu}$};
    \node[] at (2,0.5) {$f_{\mu\nu}$};
     \node[] at (-2,0.5) {$f_{\mu\nu}$};
    \node[] at (-1.8,-0.4){$F^{\rm B}_{\mu\nu}$};
    \node[] at (1.8,-0.4){$F^{\rm B}_{\mu\nu}$};
    \node[rectangle,fill]  (amp1) at (1,0) {};
    \end{tikzpicture}
    \caption{} 
    \label{Fig:feynmann_diagram_e}
\end{subfigure}
\vspace{0.5 cm}
\caption{Illustration of three types of EFT contribution to the time delay. Thick single and double wavy lines denote the background electromagnetic and gravitational fields of the shockwave, $F_{\mu\nu}^{\rm B}$ and $g_{\mu\nu}^{\rm B}$, respectively. While the EM background remains unmodified by EFT operators, we distinguish the unperturbed shockwave metric $\big(g_{\mu\nu}^{\rm B}\big)^{(0)}$ from its EFT corrections $\big(g_{\mu\nu}^{\rm B}\big)^{(c_i)}$. Thin single and double wavy lines represent the probe photon $f_{\mu\nu}$ and metric perturbation $h_{\mu\nu}$. The solid rectangle indicates interactions from EFT operators. The solid rectangle represents the interaction from EFT operators. Diagrams $(a,b)$ correspond to the first kind of contribution. Diagram $(c)$ corresponds to the second contribution. Diagrams $(d,e)$ correspond to the third contribution.
}
\label{Fig:feynmann_diagrams}
\end{figure}
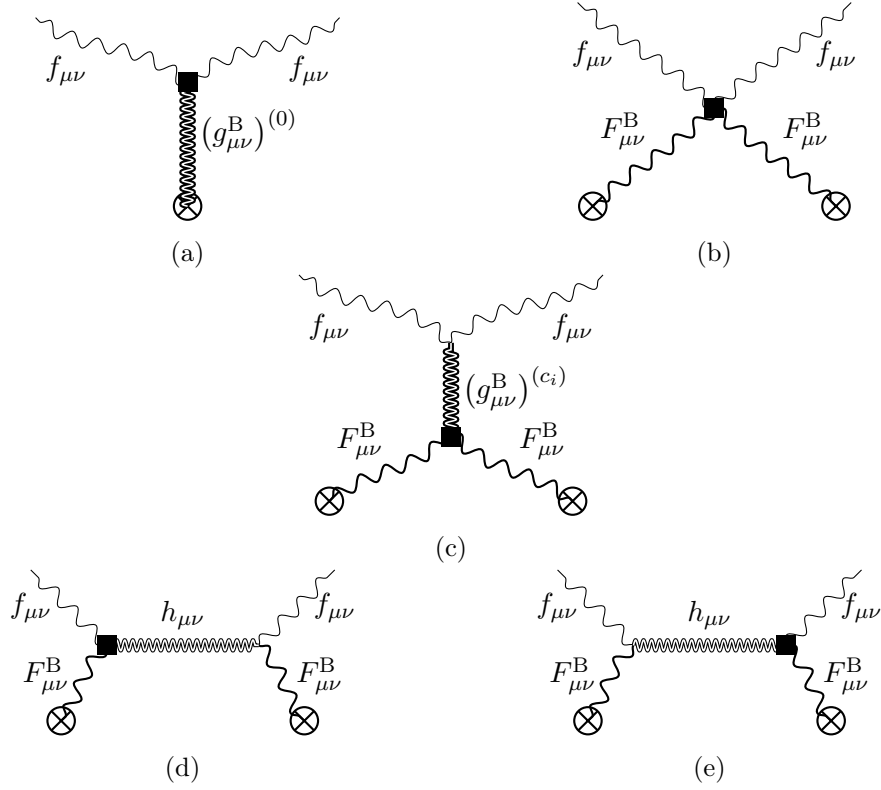

Beyond the universal contribution in Eq.~\eqref{eq:deltaV_0}, the time delay of the probe photon receives additional contributions from the EFT operators. These arise from three distinct sources: 
\begin{itemize}
 \item \textbf{First contribution: EFT induced photon-shockwave interactions.}
 
 The probe photon interacts with the background metric and EM field through the non-minimal couplings introduced by the higher derivative operators.  These contributions are illustrated with diagrams $(a,b)$ in Fig.~\ref{Fig:feynmann_diagrams}.
 \item \textbf{Second contribution: EFT corrections to the shockwave geometry.}
 
 The background metric sourced by the background EM field itself receives EFT corrections, as derived in Section~\ref{sec.3-shockwaves}. A photon minimally coupled to this corrected geometry acquires an additional time delay. This effect is represented by diagram (c).
 
    \item \textbf{Third contribution: backreaction of the probe photon.}
    
    The interaction between the probe photon and the background EM field -- either through minimal or non-minimal couplings -- induces a perturbation of the metric. In the presence of higher derivative operators, this induced perturbation generates a further contribution to the time delay proportional to the Wilson coefficients. These effects are illustrated by diagrams $(d,e)$.
\end{itemize}  
In the existing literature, only the first type of contribution was typically included, while the second and third have been overlooked. In particular, the third class of effects is often dismissed as irrelevant. From the diagrammatic viewpoint, diagrams $(d,e)$ resemble $s$-channel processes while diagrams $(a,c)$ correspond to $t$-channel exchange. Since shockwave scattering probes the eikonal limit ($s\gg t$)~\cite{Kabat:1992tb}, it is natural to assumed that 
$t$-channel effects dominate and $s$-channel contributions may be ignored. However, as we will demonstrate, this intuition does not hold in the presence of higher derivative EFT operators: the $s$-channel-like processes contain contributions that are not suppressed by an $s$-pole 
and therefore survive in the eikonal regime. As a result, both the second and third types of contributions are essential and must be included to obtain a consistent, field-redefinition–invariant expression for the total time delay. 

In the following, we derive these three classes of contributions directly from the classical equations of motion. The first two types of contributions arise solely from the equation of motion for the probe photon. To incorporate the interaction between the probe photon and the background electromagnetic field generated by the 
$F^4$ operator, we decompose the field strength as $F_{\mu\nu}=F^{\rm B}_{\mu\nu}+f_{\mu\nu}$, where $F^{\rm B}_{\mu\nu}$ is the background EM field in the shockwave and $f_{\mu\nu}$ is the perturbation associated with the probe photon. Following the procedure of Ref.~\cite{Cremonini:2023epg}, we expand the full 
EOM~\eqref{EOM-full} to first order in the perturbation. This yields
\begin{align}
    \nabla^{\rm B}_\nu f^{\mu\nu}&=2c_5\nabla^{\rm B}_\nu(R^{\rm B\,\mu\rho}f_{\rho}^{\,\nu}-R^{\rm B\,\nu\rho}f_\rho^{\,\mu})+4c_6\nabla^{\rm B}_\nu(R^{\alpha\beta\mu\nu}f_{\alpha\beta})
    \nonumber\\
    &+8c_7\nabla^{\rm B}_{\nu}\big( F^{\rm B}_{\lambda\sigma}F^{\rm{B}\,\lambda\sigma}f^{\mu\nu}+2f_{\lambda\sigma}F^{\rm B\,\lambda\sigma}F^{\rm B\,\mu\nu} \big)
    +8\tilde{c}_7\nabla_{\nu}^{\rm B}\big( \tilde{f}^{\mu\nu} F^{\rm B}_{\alpha\beta}F^{\rm{B}\,\alpha\beta}+2\tilde{F}^{\rm{B}\,\mu\nu} f_{\alpha\beta}F^{\rm{B}\,\alpha\beta} \big)
    \,.
    \label{eq:EMeom_t}
\end{align}
Here the superscript ``B'' on the covariant derivative and curvature tensors means they are evaluated with the shockwave metric. Since we are interested in the time delay at linear order in the Wilson coefficients, the left-hand side must incorporate the EFT corrections to the shockwave geometry -- i.e., we must use the EFT-corrected background metric \eqref{eq:sw-EFT}. This contribution corresponds to the second type identified earlier. The terms on the RHS arise from the corrections to the equation of motion generated by the higher derivative operators. Because these terms are already linear in 
$c_i$, their evaluation requires only the uncorrected shockwave metric \eqref{eq:sw-EM}. These constitute the first kind of contribution to the time delay. 

Considering, like in the scalar case, the probe photon propagating in the $x$-direction (see Fig.~\ref{fig: time-delay-cartesian}), we can decompose its vector potential into the radial and tangential polarizations on the transverse plane  
\begin{equation}
    A^{^{\rm probe}}
    =\phi_\rho \, d\rho + \phi_\varphi \, d\varphi
    \,.
\end{equation} Near the shockwave at 
$u=0$, the variation of the wavefunction in the transverse directions is negligible compared to its variation along $u$. Under this approximation, the 
EOM~\eqref{eq:EMeom_t} reduces to the following set of equations:
\begin{align}
   \partial_u \partial_v \phi_{\rho}+\left[\left(h(\rho)-c_5\frac{\partial_\rho h^{(0)}(\rho)}{\rho}-(c_5+4c_6)\partial^2_\rho h^{(0)}(\rho)\right)\delta(u)+32c_7 (\partial_\rho H(u,\rho))^2\right]\partial^2_v\phi_{\rho}&=0,\nonumber\\
   \partial_u \partial_v \phi_{\varphi}+\left[\left(h(\rho)-(c_5+4c_6)\frac{\partial_\rho h^{(0)}(\rho)}{\rho}-c_5\partial^2_\rho h^{(0)}(\rho)\right)\delta(u)+32\tilde{c}_7 (\partial_\rho H(u,\rho))^2\right]\partial^2_v\phi_{\phi}&=0,
   \label{eq:EOM_t}
\end{align}
  %
where $h(\rho)=h^{(0)}(\rho)+ h^{(c_i)}(\rho)$. Substituting the explicit form of $h^{(0)},  h^{(c_i)}$ and $H(u,\rho)$, and taking the $\epsilon \rightarrow 0$ limit for terms $\sim H(u,\rho)^2$, we obtain the time delay
\begin{align}
\Delta v_\rho&=-4\kappa^2 m_0\ln \rho-\frac{3\pi}{4}\frac{\kappa^2q_0^2}{\rho}-\frac{16c_6\kappa^2 m_0}{\rho^2}+\frac{\pi q_0^2}{2\rho^3}\left(3c_5\kappa^2+16c_6\kappa^2+24c_7\right),\nonumber\\
\Delta v_\varphi&=-4\kappa^2 m_0\ln \rho-\frac{3\pi}{4}\frac{\kappa^2q_0^2}{\rho}+\frac{16c_6\kappa^2 m_0}{\rho^2}+\frac{\pi q_0^2}{2\rho^3}\left(3c_5\kappa^2-2c_6\kappa^2+24\tilde{c}_7\right).
\label{eq:deltaV_t}
\end{align}
We notice that the result for $c_7$ and $\tilde{c}_7$, which receives only first type of contribution, are consistent with Ref.~\cite{Cremonini:2023epg} after accounting for a factor of two in the convention used for the black hole charge. Instead of keeping both $\mathcal{O}_5$ and $\mathcal{O}_6$, Ref.~\cite{Cremonini:2023epg} studies the operator $\alpha_3 W_{\mu\nu\rho\sigma}F^{\mu\nu}F^{\rho\sigma}$ where $W_{\mu\nu\rho\sigma}$ is the Weyl tensor.
In our basis, this choice corresponds to set $c_5\rightarrow-2\alpha_3$ and $c_6 \rightarrow \alpha_3$. 
When only the first type of contribution is included, i.e. retaining $h^{(0)}(\rho)$ alone in Eq.~\eqref{eq:EOM_t}, our results agree with Ref.~\cite{Cremonini:2023epg}. However, when the non-vanishing second type of contribution is included, our results for $W_{\mu\nu\rho\sigma}F^{\mu\nu}F^{\rho\sigma}$ obtained from the $\mathcal{O}_5$ and $\mathcal{O}_6$ operators, differ from Ref.~\cite{Cremonini:2023epg}.

A sanity check reveals that these expressions cannot be the final and physical results, because they are not invariant under the field redefinition~\eqref{eq: results-metric-redef} that shifts the Wilson coefficients as $c_5 \rightarrow c_5'=0, \; c_7\rightarrow c_7'=c_7 +\kappa^2 c_5/4 , \; \tilde{c}_7\rightarrow \tilde{c}_7'=\tilde{c}_7 +\kappa^2 c_5/{4}$. The lack of invariance signals that the calculation is missing additional EFT-induced contributions.

What is missing in the above procedure is that the electromagnetic perturbation 
$f_{\mu\nu}$ necessarily induces a perturbation of the spacetime metric as well. In other words, the backreaction of the probe photon on the geometry must be taken into account. To incorporate this effect properly, we decompose the metric as $g_{\mu \nu}=g^{\rm B}_{\mu \nu}+h_{\mu\nu}$ where $g^{\rm B}_{\mu \nu}$ is the shockwave metric and  $h_{\mu\nu}$ denotes the metric perturbation sourced by the probe. One must then expand both the full Maxwell equation \eqref{EOM-full} and the Einstein equation to first order in the perturbations and solve for $f_{\mu\nu}$ and $h_{\mu\nu}$ simultaneously. Only with this coupled system does one capture the full set of linear order EFT contributions to the time delay, including those responsible for restoring invariance under field redefinitions.

The perturbed Einstein equation takes the form\footnote{A gauge choice is required to solve this equation. In transverse–traceless gauge one has $\delta G_{\mu\nu}=-\frac{1}{2}\Box^{\rm B}  h_{\mu\nu}+R^{\rm B}_{\rho\nu\mu\sigma}h^{\rho\sigma}$.}
\begin{align}
   \kappa^2 \delta T_{\mu\nu}&=\delta G_{\mu\nu}\nonumber\\&=-\frac{1}{2}\Box^{\rm B} \bar h_{\mu\nu}+R^{\rm B}_{\rho\nu\mu\sigma}\bar h^{\rho\sigma}-\frac{1}{2}g^{\rm B}_{\mu\nu}\nabla^{\rm B}_\rho\nabla^{\rm B}_\sigma \bar h^{\rho\sigma}+\frac{1}{2}\nabla^{\rm B}_\mu\nabla^{\rm B}_\rho \bar h^{\rho\nu}+\frac{1}{2}\nabla^{\rm B}_\nu\nabla^{\rm B}_\rho \bar h^{\rho\mu},
    \label{eq:eom_hmunu}
\end{align}
where $\bar h_{\mu\nu}\equiv h_{\mu\nu}-\frac{1}{2}g^{\rm B}_{\mu\nu}\,g^{\rm B\,\rho\sigma}\,h_{\rho\sigma}$ is the trace-reversed perturbation. For the perturbed energy momentum tensor, we keep only the linear order in $f_{\mu\nu}$:
\begin{equation}
    \delta T_{\mu\nu}=\delta T^{(0)}_{\mu\nu}+ \delta T^{(c_i)}_{\mu\nu}
    = \Big[ T^{(0)}_{\mu\nu}+ T^{(c_i)}_{\mu\nu}|_{F=F^{\rm B}+f} \Big]
    \,
\end{equation}
where the bracket $[...]$ indicates retaining only terms first order in of $f_{\mu\nu}$. Since the metric perturbation receives contributions from both $\delta T^{(0)}_{\mu\nu}$ and $ \delta T^{(c_i)}_{\mu\nu}$, it is convenient to decompose $h_{\mu\nu}=h_{\mu\nu}^{(0)}+ h_{\mu\nu}^{(c_i)}$.

In the presence of the metric perturbation $h_{\mu\nu}$, the EOM for the probe photon acquires additional terms. To linear order in 
$h_{\mu\nu}$, these corrections contribute to the time delay through~\footnote{Notice that because the perturbation of field strength and metric are defined with lower indices, we have $\delta g^{\mu\nu}=-h^{\mu\nu}$ and $\delta F^{\mu\nu}=\delta(g^{\mu\alpha}g^{\nu\beta} F_{\alpha\beta})=-h^{\mu\alpha}F^{\rm B\,\nu}_{\alpha}-h^{\nu\beta} F^{\rm B\,\mu}_{\;\;\;\;\beta}+ f^{\mu\nu}$.}
\begin{align}
    \partial_\nu f^{\mu\nu}
    &+\delta \Gamma^{\mu}_{\;\nu\lambda} F^{\rm B\,\lambda\nu}
    +\delta\Gamma^\nu_{\,\nu\lambda}F^{\rm B\,\mu\lambda}-\nabla^{\rm B}_\nu(h^{\mu\lambda}F_\lambda^{\rm B\, \nu})-\nabla^{\rm B}_\nu(h^{\nu\lambda}F_{\;\;\;\;\lambda}^{\rm B\, \mu})
    \nonumber\\
    &=2c_5\nabla^{\rm B}_\nu \Big(\delta (R^{\mu\rho})F_{\rho}^{\rm B\,\nu}-\delta(R^{\nu\rho})F_\rho^{\rm B\,\mu} \Big)
    +4c_6\nabla^{\rm B}_\nu \Big(\delta(R^{\alpha\beta\mu\nu})F^{\rm B}_{\alpha\beta}\Big)
    \,.
\label{eq:EMeom_s}
\end{align}
On the LHS, the perturbed Christoffel symbols take form 
\begin{equation}
    \delta \Gamma^\lambda_{\,\mu\nu}=\frac{1}{2}g^{\lambda\rho} \big( \nabla^B_\mu h_{\rho\nu}+\nabla^B_\nu h_{\rho\mu}-\nabla^B_\rho h_{\mu\nu} \big)
    \,.
\end{equation}
If we substitute $h_{\mu\nu}^{(0)}$ into the LHS of photon EoM~\eqref{eq:EMeom_s}, it may appear that, even in the absence of effective operators, the $s$-channel–like contribution could modify the universal time delay $\Delta v^{(0)}$ in Eq.~\eqref{eq:deltaV_0}. However, this does not occur: in pure Einstein--Maxwell theory, the eikonal limit applies to shockwave scattering, and such $s$-channel effects are negligible. The reason can be understood as follows. In general, the metric perturbation obtained from Eq.~\eqref{eq:eom_hmunu} contains the propagator in the shockwave background and has the schematic form: $h_{\mu\nu}\sim \frac{1}{\partial_u\partial_v+h\delta(u)\partial_v^2}\delta T_{\mu\nu}$. Near the shockwave at $u=0$,  $h_{\mu\nu}^{(0)}$ is highly suppressed by $\delta(u)$ and does not need to be included in the calculation of $\delta \Gamma^\lambda_{\,\mu\nu}$, and the corresponding $s$-channel-like terms do not enter the physical time delay.

In the presence of higher derivative EFT operators, however, two effects change this conclusion. First, the higher derivative structure of $ \delta T_{\mu\nu}^{(c_i)}$ can partially cancel the graviton propagator appearing in the solution for $h_{\mu\nu}$, producing a \textit{local} contribution to $h_{\mu\nu}^{(c_i)}(x)$. By \textit{local} we mean a term that depends only on the values of the probe and background fields and their derivatives evaluated at the same spacetime point $x$.  We will demonstrate this explicitly for the operator $\mathcal{O}_5$ (the contribution from the operator $\mathcal{O}_6$ is more cumbersome to compute and will be reported in a forthcoming work~\cite{Grojean:2026} and therefore for now on $c_6=0$). Second, although 
$h_{\mu\nu}^{(0)}$ remains suppressed, it can nevertheless generate finite and local contributions to $\delta R_{\mu\nu\rho\sigma}$ and $\delta R_{\mu\nu}$, which appear on the RHS of Eq.~\eqref{eq:EMeom_s} when effective operators are present. Because we are working only to linear order in $c_i$, the RHS requires only the contributions from $h_{\mu\nu}^{(0)}$. This is why we must account for perturbed Riemann and Ricci tensors on the RHS, but not perturbed Christoffel symbols inside the covariant derivatives.

The full solution for $h_{\mu\nu}$ in the presence of higher derivative operators -- i.e. solving Eq.~\eqref{eq:eom_hmunu} exactly -- requires substantial additional work, and we defer this analysis to our upcoming paper~\cite{Grojean:2026}. Here, however, we show that in order to capture the relevant $s$-channel–like contribution from the 
$\mathcal{O}_5$ operator, it is not necessary to solve explicitly for 
$h_{\mu\nu}$ from Eq.~\eqref{eq:eom_hmunu}. Instead, we demonstrate that the required contribution can be obtained directly from the structure of the perturbation equations. Once this additional term is included, the resulting time delay becomes fully invariant under field redefinitions, as expected for any physical observable in EFT.

First, observe that on the RHS of Eq.~\eqref{eq:EMeom_s} the perturbations of the Ricci tensor can be expressed using the Einstein equations. Keeping only terms linear in the electromagnetic fluctuation 
$f_{\mu\nu}$, we obtain
\begin{equation}   \delta(R_{\mu\nu})=\kappa^2\delta(T^{(0)}_{\mu\nu}-\frac{1}{2}g_{\mu\nu}T^{(0)})=\kappa^2 \Big[ (F_{\mu\rho}F^{\nu\rho}-\frac{1}{4}g_{\mu\nu}F^2)|_{F=F^{\rm B}+f} \Big]
\,.
\end{equation}

Next, consider the LHS of Eq.~\eqref{eq:EMeom_s}. The key point is that the metric perturbation 
$ h_{\mu\nu}^{(c_5)}$, sourced by the 
$\mathcal{O}_5$ operator, is directly related to the field-redefinition shift 
$\delta g^{(c_5)}_{\mu\nu}$. To see this explicitly, rewrite the Einstein equations schematically as
\begin{equation}
    \int dy^4 \frac{\delta \big(\sqrt{-g}\, R/2\kappa^2 \big)_y}{\delta g_{\mu\nu}(x) }+ \int dz^4 \frac{\delta \big(c_5\sqrt{-g}\, R_{\mu\nu}F^{\mu\rho}F^{\nu}_{\;\rho} \big)_z}{\delta g_{\mu\nu}(x) }=0,
\end{equation}
where the subscripts $y,z$ indicate the space-time points. Perturbing this equation yields 
\begin{equation}
    \int dz^4 dy^4\frac{\delta^2 \big(\sqrt{-g}\, R/2\kappa^2 \big)_y}{\delta g_{\alpha\beta}(z)\delta g_{\mu\nu}(x) } h^{(c_5)}_{\alpha \beta}(z)+ \int dz^4 \frac{\delta \Big(c_5\sqrt{-g}\,R_{\mu\nu}\big[F^{\mu\rho}F^{\nu}_{\;\rho}|_{F=F^{\rm B}+f} \big] \Big)_z}{\delta g_{\mu\nu}(x) }=0
    \,,
\end{equation}
On the other hand, the field redefinition used in Eq.~\eqref{eq: shifted-action-c5} satisfies
\begin{equation}
    \int dy^4 \frac{\delta \big(\sqrt{-g}\, R/2\kappa^2 \big)_y}{\delta g_{\mu\nu}(x) }\delta g^{(c_5)}_{\mu\nu}(x) 
    + c_5\big(\sqrt{-g}\,R_{\mu\nu}F^{\mu\rho}F^{\nu} \big)_x=0
    \,.
\end{equation} 
Comparing the two relations immediately gives
\begin{equation}
     h_{\mu\nu}^{(c_5)}=\Big[\delta g^{(c_5)}_{\mu\nu}|_{F=F^{\rm B}+f}\Big]
     =2\kappa^2c_5 \Big[(F_\mu^\rho F_{\nu\rho}-\frac{1}{2}g_{\mu\nu}F_{\mu\nu}F^{\mu\nu})|_{F=F^{\rm B}+f} \Big]
     \,. 
\end{equation}

With the explicit expressions for $\delta R_{\mu\nu}$ and $ h_{\mu\nu}^{(c_5)}$, we substitute them into Eq.~\eqref{eq:EMeom_s} and extract the resulting 
$s$-channel-like contribution to the time delay. For both polarizations we obtain:
\begin{align}
\Delta v_\rho^{s(c_5)}=\Delta v_\varphi^{s(c_5)}=\frac{3\pi \kappa^2q_0^2}{2\rho^3}c_5.
\end{align}
After adding this contribution to Eq.~\eqref{eq:deltaV_t}, the full contribution from operators $\{\mathcal{O}_5,\mathcal{O}_7,\tilde{\mathcal{O}}_7\}$ to time delay becomes 
\begin{align}
\Delta v_\rho&\supset\frac{\pi q_0^2}{\rho^3}(3c_5\kappa^2+12c_7),\nonumber\\
\Delta v_\varphi&\supset\frac{\pi q_0^2}{\rho^3}(3c_5\kappa^2+12\tilde{c}_7),
\end{align}
which are invariant under the field redefinition. 

At this stage one might wonder whether the appearance of an 
$s$-channel–like contribution is merely an artifact of the operator $\mathcal{O}_5$ which can be fully removed with redefinition of the metric, and an operator like $\mathcal{O}_6$ 
might not generate an analogous effect. However, this is unlikely. The essential distinction is that the contributions from $\mathcal{O}_5$ -- whether through $ h^{(c_5)}_{\mu\nu}$ on the LHS of Eq.~\eqref{eq:EMeom_s} or through $\delta R_{\mu\nu}$ on the RHS -- are purely local, reflecting the fact that $\mathcal{O}_5$ can be removed with local field redefinition. By contrast, $\mathcal{O}_6$ can not be removed with redefinition of the metric, so neither $ h^{(c_6)}_{\mu\nu}$ nor the curvature perturbation $\delta R_{\mu\nu\rho\sigma}$ is expected to be purely local; both generally involve pieces with graviton propagators. Nevertheless, in both cases local pieces can still arise: the higher derivatives in $\mathcal{O}_6$ can partially cancel the propagator, producing local/contact contributions that are not suppressed in the eikonal limit and therefore contribute to the time delay. A natural conjecture is that for certain linear combinations of $\mathcal{O}_5$ and $\mathcal{O}_6$, for instance in the form of $W_{\mu\nu\rho\sigma}F^{\mu\nu}F^{\rho\sigma}$, the local pieces might cancel, leading to a vanishing $s$-channel-like contribution. Verifying this, however, requires an explicit solution of the perturbed Einstein equation, and we leave this analysis to future work~\cite{Grojean:2026}.

\section{Conclusion and Outlook}
\label{sec.5-conc-outl}
In this work, we have studied shockwaves arising from ultra-relativistic boosts of electrically charged black holes in Einstein--Maxwell effective field theory. Working to linear order in the Wilson coefficients, we derived the EFT corrections to the charged shockwave geometry and computed the time delay experienced by a probe photon traversing it. A key finding is that two contributions neglected in previous analyses are essential for obtaining a physically meaningful result: the EFT corrections to the shockwave metric itself, and the metric perturbation sourced by the interaction of the probe photon with the electromagnetic shockwave background. Only when both effects are included does the total time delay remain invariant under field redefinitions of the metric, as required for consistency within the EFT framework. 

There are actually cases where the backreaction effects can bring the first non-vanishing contributions to the time delay generated by certain EFT operators.
For instance, in the 5D Einstein--Maxwell EFT, the $R_{\mu\nu\rho\sigma}R^{\mu\nu\rho\sigma}$ operator can not be removed in the minimal operator basis because the Gauss--Bonnet term is no longer a boundary term. While this operator is not probed by the photon equation of motion on the unperturbed shockwave metric~\cite{Cremonini:2024lxn}, its contribution through the perturbative corrections to the
shockwave metric (i.e. the backreaction effect) induces a dependence of the full time delay.

It would also be interesting to consider the effects of not rescaling the black hole mass and charge with the boost parameter: as shown in Ref.~\cite{Cremonini:2024lxn}, this would allow one to study the near-horizon regime of extremal black holes. We leave the study of this different limit for future work.

An immediate next step is to compute the remaining 
$s$-channel contribution from 
$\mathcal{O}_6$, which requires solving the perturbed Einstein equations on the shockwave background. Completing this calculation will provide the full EFT-corrected time delay for charged shockwaves in Einstein--Maxwell theory.

It is also instructive to compare our results with those obtained from on-shell amplitude methods. For purely gravitational shockwaves, Ref.~\cite{Camanho:2014apa} demonstrated that the classical time delay can be extracted by cutting the tree-level four-point amplitude. In the presence of both electromagnetic and gravitational fields, however, the amplitude structure becomes richer: the relevant discontinuities arise from cuts of loop-level amplitudes, as illustrated in the analysis of Ref.~\cite{AccettulliHuber:2020oou}. Matching our shockwave calculation with this amplitude-based approach would provide a non-trivial cross-check of how classical observables emerge from EFT scattering amplitudes.

With the complete expression for the time delay in hand, one can systematically explore the resulting causality bounds on the Einstein--Maxwell EFT. These bounds offer a complementary probe of the UV consistency of gravity--gauge systems and may shed further light on conjectures such as weak gravity.

\subsubsection*{Acknowledgements} 
This work was funded by the Deutsche Forschungsgemeinschaft under Germany’s Excellence Strategy -- EXC 2121 “Quantum Universe” -- 390833306. This project also has received funding from the European Union’s Horizon Europe research and innovation programme under the Marie Sk\l odowska-Curie Staff Exchange grant agreement No 101086085 -- ASYMMETRY. Many manipulations were
carried out with the help of the Mathematica package \texttt{xAct}~\cite{xAct}.

\addcontentsline{toc}{section}{References}
\bibliographystyle{JHEP}
\bibliography{main.bib}
\end{document}